\documentclass[aps,prd,preprint,groupedaddress,showpacs,11pt]{revtex4}

\usepackage{amsmath}
\usepackage{amssymb}
\usepackage{graphics}
\usepackage{graphicx}
\usepackage{psfrag}
\usepackage[mathcal]{eucal}

\def\gsim{\buildrel > \over {_{\sim}}}

\begin{document}

\preprint{FERMILAB-Pub-08-038-T}
\preprint{UCI-TR-2008-8}

%Title of paper

\title{Lepton flavor violation in predictive supersymmetric GUT models}

\author{Carl H. Albright$^{1,2,}$}
\email{albright@fnal.gov}
\author{Mu-Chun Chen$^{3,}$}
\email{muchunc@uci.edu}
%\homepage[]{Your web page}
%\thanks{}
%\altaffiliation{}
\affiliation{$^1$Department of Physics, Northern Illinois University, DeKalb,
  IL 60115\\
$^2$Fermi National Accelerator Laboratory, Batavia, IL 60510\\
$^3$Department of Physics \& Astronomy,~University of California, Irvine,
  CA 92697}
\date{March 12, 2008\\[0.75in]}
%\date{\today}

\begin{abstract}
There have been many theoretical models constructed which aim to explain the neutrino 
masses and mixing patterns. While many of the models will be eliminated once more 
accurate determinations of the mixing parameters, especially $\sin^2 2\theta_{13}$,
are obtained, charged lepton flavor violation experiments are able to differentiate 
even further among the models. In this paper, we investigate various rare lepton flavor 
violation processes, such as $\ell_{i} \rightarrow \ell_{j} + \gamma$ and $\mu-e$ 
conversion, in five predictive supersymmetric (SUSY) $SO(10)$ models and their 
allowed soft-SUSY breaking parameter space in the constrained minimal SUSY standard 
model. Utilizing the Wilkinson Microwave Anisotropy Probe dark matter constraints, we 
obtain lower bounds on the branching ratios of these rare processes and find that at 
least three of the five models we consider give rise to predictions for 
$\mu \rightarrow e + \gamma$ that will be tested by the MEG Collaboration at PSI.  In 
addition, the next generation $\mu-e$ conversion experiment has sensitivity to the 
predictions of all five models, making it an even more robust way to test these 
models. While generic studies have emphasized the dependence of the branching ratios 
of these rare processes on the reactor neutrino angle, $\theta_{13}$, and the mass of 
the heaviest right-handed neutrino, $M_3$, we find very massive $M_3$ is more 
significant than large $\theta_{13}$ in leading to branching ratios near to the 
present upper limits.
\end{abstract}

% insert suggested PACS numbers in braces on next line
% insert suggested keywords - APS authors don't need to do this
%\keywords{}

\vspace*{0.5in}

%\maketitle must follow title, authors, abstract, \pacs, and \keywords
\maketitle

% body of paper here - Use proper section commands
% References should be done using the \cite, \ref, and \label commands

\section{Introduction}

The recent compilation of oscillation data from the atmospheric \cite{atm}, reactor 
\cite{reactor}, and long baseline \cite{lbl} neutrino experiments has provided 
solid evidence that neutrinos have small but non-zero masses. A global fit to current 
data gives the following $2\sigma$ limits for the mixing 
parameters~\cite{Maltoni:2004ei},
\begin{eqnarray}
\sin^{2} \theta_{12} & = &  (0.28 - 0.37), \quad \Delta m_{21}^{2} =  (7.3 - 8.1)  
\times 10^{-5} 
	\; \mbox{eV}^2\\
\sin^{2} \theta_{23} & = &  0.38 - 0.63, \quad\quad \Delta m_{31}^{2} = (2.1 - 2.7) 
\times 10^{-3}
	 \; \mbox{eV}^2\\
\sin^{2} \theta_{13} & < &  0.033 \; .
\label{eq:data}
\end{eqnarray}

\noindent Since then, the measurements of neutrino oscillation parameters have entered 
a precision era. On the other hand, as no information exists for the value of 
$\theta_{13}$, the Dirac or Majorana nature of the neutrinos, the Dirac and/or Majorana 
CP phases, and the neutrino mass hierarchy, there are discoveries that are still yet to 
come.   

In the Standard Model (SM), due to the lack of  right-handed neutrinos and the 
conservation of lepton numbers, neutrinos are massless. To generate non-zero 
neutrino masses thus calls for physics beyond the SM. There have been many theoretical 
ideas proposed with an attempt to accommodate the experimentally observed small neutrino 
masses and the larger mixing angles among them. In Ref.~\cite{Albright:2006cw}, we have 
surveyed 63 models in the literature that are still viable candidates and have 
reasonably well-defined predictions for $\theta_{13}$.  We found that the predictions 
for $\sin^{2} 2 \theta_{13}$ of half of the models cover the range from 0.015 to the 
present upper bound of 0.13. Consequently, half of the models can be eliminated in the 
next generation of reactor experiments. 

One of the implications of the observation of neutrino oscillation is the possibility 
of measurable branching ratio for charged lepton flavor-violating (LFV) decays. While
not the case in the SM, predictions of the supersymmetric (SUSY) Grand Unified Theories 
(GUT) for these rare decays are much enhanced, as these processes are suppressed by the 
SUSY scale, rather than the Plank scale \cite{BorzMas}. Furthermore, as 
different models obtain large neutrino mixing angles through different mechanisms, their 
predictions for the LFV charged lepton decays can be very distinct. Consequently, LFV 
charged lepton decays may provide a way to distinguish different SUSY GUT models.

Among the models aiming to explain the neutrino masses and mixing, a particularly 
promising class are those based on (SUSY) SO(10); for recent reviews of SO(10) models, 
see \cite{Chen:2003zv}. In this paper, we investigate the predictions for various LFV 
charged lepton decays as well as muon-electron conversion in five of the SUSY SO(10) 
models, assuming constrained minimal SUSY standard model (CMSSM) boundary conditions 
where only five soft-SUSY breaking parameters are present. Furthermore, we impose the 
Wilkinson Microwave Anisotropy Probe (WMAP) dark matter constraints in the neutralino, 
stau and stop coannihilation regions. Specifically, we present the allowed values for 
these soft-SUSY parameters for various branching ratios of these rare LFV processes. 
In addition, the lower bounds on the predictions for these rare processes 
in the five different SUSY SO(10) models are given. We find that the 
predictions in these models are very distinct. We note the crucial role of the WMAP 
constraints in deducing the lower bounds on the predictions. 

Many authors have previously studied the branching ratio predictions for charged
LFV decays in the SUSY GUT framework.  Rather than study specific models, they 
generally adopt a generic approach and assume a nearly diagonal 
Cabibbo-Kobayashi-Maskawa (CKM)-like or bimaximal Pontecorvo-Maki-Nakagawa-Sakata 
(PMNS)-like mixing matrix to diagonalize the Yukawa neutrino matrix \cite{generic}.
Following the procedure of Casas and Ibarra \cite{CasasIbarra} to invert the seesaw 
formula, they have carried out Monte Carlo studies by scanning the unknown right-handed 
neutrino mass spectrum and the angles and phases of the inversion matrix in order to 
present scatter plots of the rare branching ratios.  A few exceptions to this procedure 
can be found in Ref. \cite{exceptions}.  Here we are interested in specific models in 
order to determine the ability of the LFV experiments to differentiate among and rule 
out some models.  The models chosen are highly predictive and illustrate a wide variety 
of textures for the charged lepton and neutrino mass matrices, and predictions for the
right-handed neutrino mass spectrum and the reactor neutrino mixing angle, $\theta_{13}$.

In Sec. II we describe the five representative SUSY SO(10) models that we have 
analyzed.  In Sec. III we review LFV charged lepton decays in the SM and SUSY GUTs and 
present the predictions for the five SUSY SO(10) models considered.  In Sec. IV their 
expectations for $\mu-e$ conversion are given. Sect. V concludes this paper. 

\section{PREDICTIVE SUPERSYMMETRIC GRAND UNIFIED MODELS CONSIDERED}

We begin with a brief discussion of the general formalism on which the 
supersymmetric $SO(10)$ grand unified models are based.  For all five 
models to be illustrated, the seesaw mechanism \cite{seesaw} is of the conventional 
type I leading to normal hierarchies for the light Majorana 
neutrinos.  The leptonic sector of the Yukawa superpotential at the GUT 
scale can then be written as 
\begin{equation}
  W_Y = N^c_i Y^\nu_{ij} L_j H_u + E^c_i Y^e_{ij} L_j H_d 
           + N^c_i M_{Rij} N^c_j,
  \label{eq:Yuk}
\end{equation}

\noindent where $L$ represents the left-handed lepton doublet, $N^c$ the 
left-handed conjugate neutrino singlet, and $E^c$ the left-handed conjugate
charged lepton of one of the three ${\bf 16}$ dimensional representations of 
$SO(10)$.  When the two Higgs doublets, $H_u$ and $H_d$, acquire vacuum
expectation values, the charged lepton and neutrino mass matrices are 
generated and can be written in $6 \times 6$ form \\ 
\begin{equation}
\begin{array}{rcl}
  {\cal M}^e &=& \left(\begin{array}{cc} e^T_L & e^{cT}_L\\ \end{array}\right)
    \left(\begin{array}{cc}
      0 & M^T_E\\ M_E & 0\\ \end{array}\right)
    \left(\begin{array}{c}
      e_L\\ e^c_L\\ \end{array}\right),\\[0.1in]
  {\cal M}^\nu &=& \left(\begin{array}{cc} 
    \nu^T_L & N^{cT}_L\\ \end{array}\right)
    \left(\begin{array}{cc}
      0 & M^T_N\\ M_N & M_R\\ \end{array}\right)
    \left(\begin{array}{c}
      \nu_L \\ N^c_L\\ \end{array}\right).
\end{array}
\label{eq:massmat}
\end{equation}

\noindent  With the entries in $M_R$ much larger than those in $M_N$, the
light Majorana neutrino mass matrix is given by the well-known
type I seesaw formula \cite{seesaw},
\begin{equation}
  M_\nu = - M^T_N M^{-1}_R M_N.
  \label{eq:seesaw}
\end{equation}

Note that the above Dirac $3 \times 3$ mass matrix entries, $M_E$ and 
$M_N$, are written in right - left order and appear in the $SO(10)$ flavor
basis.  As such, the matrices can be diagonalized by the unitary 
transformations
\begin{equation}
  \begin{array}{rcl}
  U^\dagger _{ER} M_E U_{EL} &=& {\rm diag}(m_e,\ m_\mu,\ m_\tau),\\
  U^\dagger _{MR} M_R U^*_{MR} &=& {\rm diag}(M_1,\ M_2,\ M_3),\\
  U^T_{\nu L} M_\nu U_{\nu L} &=& {\rm diag}(m_1,\ m_2,\ m_3),\\
  \end{array}
  \label{eq:diag}
\end{equation}

\noindent by forming the Hermitian products, calculating the left and right
unitary transformations of column eigenvectors, and phase rotating the 
complex symmetric $M_R$ and $M_\nu$ matrices, so their mass eigenvalues 
are real.  The PMNS neutrino mixing matrix \cite{PMNS} is then given by 
$V_{PMNS} = U^\dagger _{EL} U_{\nu L}$.

On the other hand for later use, it is convenient to transform to the bases 
where $M_E$ and $M_R$ are diagonal and denoted by primed quantities.
In this case 
\begin{equation}
  \begin{array}{rcl}
  M'_\nu = - M_N'^T M_R'^{-1} M'_N,\\
  U'^T _{\nu L} M'_\nu U'_{\nu L} = {\rm diag}(m_1,\ m_2,\ m_3),\\
  \end{array}
  \label{eq:diagbas}
\end{equation}

\noindent where now $V_{PMNS} = U'_{\nu L}$, since $U'_{EL}$ is just the 
identity matrix.  A comparison of the seesaw
diagonalization matrices then reveals that the transformed Dirac neutrino 
matrix $M'_N$ in the new basis is given by
\begin{equation}
   M'_N = U^\dagger _{MR} M_N U_{EL},
\label{eq:MN'}
\end{equation}

\noindent in terms of the original matrix in the flavor bases.

With either basis, the left-handed neutrino PMNS mixing matrix can be 
written as $V_{PMNS} = U_{PMNS}\Phi$, where by convention  \cite{PDB}
\begin{equation}
  U_{PMNS} = \left(
  \begin{array}{ccc}
   c_{12}c_{13} & s_{12}c_{13} & s_{13}e^{-i\delta}\\
       -s_{12}c_{23} - c_{12}s_{23}s_{13}e^{i\delta} & 
       c_{12}c_{23} - s_{12}s_{23}s_{13}e^{i\delta} & s_{23}c_{13} \\
       s_{23}s_{12} - c_{12}c_{23}s_{13}e^{i\delta} &
       -c_{12}s_{23} - s_{12}c_{23}s_{13}e^{i\delta} & c_{23}c_{13} 
       \end{array}\right)
\label{eq:UPMNS}
\end{equation}

\noindent in terms of the three mixing angles, $\theta_{12},\ \theta_{23}$ and 
$\theta_{13}$; and the Dirac $CP$ phase, $\delta$, in analogy with the quark 
mixing matrix. The Majorana phase matrix, 
\begin{equation}
  \Phi = {\rm diag}(e^{i\chi_1},\ e^{i\chi_2},\ 1),
\label{eq:Phi}
\end{equation}

\noindent written in terms of the two Majorana phases, $\chi_1$ and $\chi_2$,
is required since an arbitrary phase transformation is 
not possible when one demands real diagonal neutrino mass entries in the 
transformation of $M_\nu$ in Eq. (\ref{eq:diag}) or $M'_\nu$ in 
Eq. (\ref{eq:diagbas}). 

With this background in mind, we now turn to a brief discussion of the five 
$SO(10)$ models we have considered for our lepton flavor violation study.  
The $SO(10)$ grand unification symmetry is an economical and attractive one
\cite{Chen:2003zv}, for all sixteen left-handed quark and lepton fields and 
their left-handed conjugates fit neatly into one ${\bf 16}$ representation per 
family.  Many models exist in the literature which differ from one another by 
their Higgs representation assignments and type of flavor symmetry imposed, if any.  
To appreciate this, it is of interest to note the following decompositions of 
the direct product of representations:
\begin{equation}
\begin{array}{rcl}
  {\bf 16 \otimes 16} &=& {\bf 10_s \oplus 120_a \oplus 126_s}, \\
  {\bf 16 \otimes \overline{16}} &=& {\bf 1 \oplus 45 \oplus 210}, \\
\end{array}
\label{products}
\end{equation}

\noindent where in the first product the ${\bf 10}$ and ${\bf 126}$ matrices
are symmetric, while the ${\bf 120}$ is antisymmetric under the interchange of 
family indices.

Given this group structure for $SO(10)$, there are two general classes of 
models which have been extensively studied.  Those with Higgs in the ${\bf 10,
\ 126,\ \overline{126}}$, and possibly the ${\bf 120}$ and/or ${\bf 210}$,
dimensional representations are often referred to as the minimal Higgs 
models \cite{minimal} and lead to symmetric or antisymmetric matrix elements, or a 
superposition of the two.  The advantage is that the couplings are 
renormalizable and preserve R-parity, but the latter representations are of 
rather high rank and disfavored in the string theory framework.  The other 
class typically involves the lower rank Higgs representations such as 
${\bf 10,\ 16,\ \overline{16}, 45}$, with some non-renormalizable effective 
operators formed from them, \cite{lopsided} but R-parity is not conserved.  
They can lead to lopsided mass matrices for the charged lepton and down quark 
mass matrices, due to the $SU(5)$ structure present in the ${\bf 16}$'s.  As such, 
one may anticipate that they will predict a higher level of lepton flavor 
violation than the first class.  
The $SU(2)_{L}$ triplet components in the Higgs representations in 
all five models considered have no electroweak VEV 
and hence lead to the conventional type I seesaw
mechanism with the prediction of normal hierarchy for the light left-handed
neutrino spectrum \cite{normal}.  We elaborate on each model in turn but give only the 
neutrino and charged lepton mass matrices.  All successfully predict the 
observed quark structure and CKM mixings.

\subsection{Albright - Barr $SO(10)$ Model}

This model, based on $SO(10)$ with a $U(1) \times Z_2 \times Z_2$ flavor 
symmetry  \cite{ABmodel}, is of the lopsided variety and has matrices of the following 
textures:
\begin{equation}
  M_N = \left(\begin{array}{ccc} \eta & \delta_N & \delta'_N \\
          \delta_N & 0 & - \epsilon\\ \delta'_N & \epsilon & 1\\ 
          \end{array}\right) m_U,
  \qquad M_E = \left(\begin{array}{ccc} 0 & \delta & \delta'\\
          \delta & 0 & - \epsilon\\ \delta' & \sigma + \epsilon & 1\\ 
          \end{array}\right) m_D,
  \label{eq:MNMEAB}
\end{equation}

\noindent where the parameters have the values $\eta = 1.1 \times 10^{-5},\ 
\epsilon = 0.147,\ \sigma = 1.83,\ \delta = 0.00946 ,\ 
\delta' = 0.00827 e^{i119.4^\circ},\ \delta_N = -1.0\times 10^{-5},\ 
\delta'_N = -1.5\times 10^{-5},\ m_U = 113\ {\rm GeV},\ m_D = 1\ {\rm GeV}$.  
The right-handed Majorana mass matrix is given by 
\begin{equation}
  M_R = \left(\begin{array}{ccc} c^2 \eta^2 & -b \epsilon \eta & a\eta\\
         -b\epsilon \eta & \epsilon^2 & -\epsilon\\
         a\eta & -\epsilon & 1\\ \end{array}\right) \Lambda_R,
  \label{eq:MRAB}
\end{equation}

\noindent where $a = c = 0.5828i,\ b = 1.7670i,\ \Lambda_R = 2.35 \times 
10^{14}$ GeV.  One finds 
\begin{equation}
  \begin{array}{rl}
      &M_1 \simeq M_2 \simeq 4.45\times 10^8\ {\rm GeV},\quad 
      M_3 = 2.4 \times 10^{14}\ {\rm GeV},\\
      &m_1 = 3.11\ {\rm meV},\quad m_2 = 9.48\ {\rm meV},\quad m_3 = 49.13\ 
           {\rm meV},\\
      &\Delta m^2_{21} = 8.0 \times 10^{-5}\ {\rm eV^2},\quad  
          \Delta m^2_{32} = 2.3 \times 10^{-3}\ {\rm eV^2},\\
      &\sin^2 2\theta_{23} = 0.99,\quad \sin^2 \theta_{12} = 0.28,\quad 
          \sin^2 \theta_{13} = 0.0020.\\
  \end{array}
\label{eq:ABresults}
\end{equation}

\noindent  A value of $\tan \beta = 5$ assures that $\sigma \gg \epsilon$
and that the corresponding lopsided nature of the 23 element of the down 
quark mass matrix gives a good fit to the results for the quark sector.
The $\delta_N,\ \delta'_N$ elements of the Dirac neutrino matrix were added 
\cite{Amodel} to the original model to give a better fit to baryogenesis 
arising from resonant leptogenesis involving the two lighter right-handed 
neutrinos \cite{leptogen}.
Evolution downward from the GUT scale to $M_Z$ has little effect on the 
mixing angles due to the small $\tan \beta$, the opposite CP-parity of the 
lighter two right-handed Majorana neutrinos, and their large hierarchy with
the heaviest one; but serves to lower the values of the $m_i$'s and 
$\Delta m^2_{32}$ and $\Delta m^2_{21}$ relative to their GUT scale values.

\subsection{Chen - Mahanthappa $SO(10)$ Model}

This model, based on $SO(10)$ with a $SU(2) \times Z_2 \times Z_2 \times Z_2$
 flavor symmetry \cite{CMmodel}, is of the minimal Higgs variety leading
to symmetric entries in the mass matrices:
\begin{equation}
  M_N = \left(\begin{array}{ccc} 0 & 0 & a\\ 0 & be^{i\theta} & c\\ 
          a & c & 1\\ \end{array}\right) vd\sin \beta,
  \qquad M_E = \left(\begin{array}{ccc} 0 & e e^{-i\phi} & 0\\
          e e^{i\phi} & -3f & 0\\ 0 & 0 & 1\\ 
          \end{array}\right) vh\cos \beta,
  \label{eq:MNMLCM}
\end{equation}

\noindent where $v = 174$ GeV and the parameters have the values 
$a = 0.00250,\ b = 0.00326,\ c = 0.0346,\ d = 0.650,\ e = 0.004036,
\ f = 0.0195,\ h = 0.06878,\ \theta = 0.74,\ \phi = -1.52$.  
The solar and atmospheric neutrino mass squared differences,
$\Delta m^2_{\rm sol} = 8.14 \times 10^{-5}\ {\rm eV^2}$ and 
$\Delta m^2_{32} = 2.3 \times 10^{-3}\ {\rm eV^2}$ were used as input to 
determine the $t = 0.344$ and $M_3 = 6.97 \times 10^{12}$ parameters in the 
effective light left-handed Majorana neutrino mass matrix 
\begin{equation}
  M_{\nu_L} = \left(\begin{array}{ccc} 0 & 0 & t\\ 0 & 1 & 1 + t^n\\
         t & 1 + t^n & 1\\ \end{array}\right) \frac{(vd \sin \beta)^2}{M_R},
  \label{eq:MLCM}
\end{equation}

\noindent  One then finds 
\begin{equation}
  \begin{array}{rl}
      &M_1 = 1.09 \times 10^7\ {\rm GeV},\quad M_2 = 4.53 \times 10^9\ 
           {\rm GeV},\quad M_3 = 6.97 \times 10^{12}\ {\rm GeV},\\
      &m_1 = 2.62\ {\rm meV},\quad m_2 = 9.39\ {\rm meV},\quad m_3 = 49.2\ 
           {\rm meV}.\\
      &\sin^2 2\theta_{23} = 1.00,\quad \sin^2 \theta_{12} = 0.27,\quad 
          \sin^2 \theta_{13} = 0.013.\\
  \end{array}
\label{eq:CMresults}
\end{equation}
 
\noindent  For this model, a value of $\tan \beta = 10$ is used.  The effect
of evolution from the GUT scale to $M_Z$ is to raise $\sin^2 \theta_{12}$  
and to lower $\sin^2 \theta_{13}$.

\subsection{Cai - Yu $SO(10)$ Model}

This model is based on $SO(10)$ with an $S_4$ flavor symmetry  \cite{CYmodel}.  
The Higgs fields appear in six ${\bf 10}$'s, three $\overline{\bf 126}$'s, three
${\bf 126}$, and one ${\bf 210}$ representations of $SO(10)$, distinguished 
by their $S_4$ flavor assignments.  Of the 14 pairs of Higgs doublets at the 
GUT scale, all but one pair are assumed to get superheavy.  The charged 
lepton mass matrix is chosen to be diagonal, 
while the right-handed Majorana neutrino mass matrix is proportional to 
the identity matrix, so all three heavy neutrinos are degenerate.  The 
Dirac neutrino and charged lepton mass matrices are symmetric and given by
\begin{equation}
\begin{array}{rcl}
  M_N &=& \left(\begin{array}{ccc} a_0 - 2 a_2 - 3(d_0 - 2 d_2) & a_5 & a_4\\
          a_5 & a_0 + a_1 + a_2 -3(d_0 + d_1 +d_2) & a_3\\
	a_4 & a_3 & a_0 - a_1 + a_2 - 3(d_0 - d_1 + d_2)\\ 
          \end{array}\right),\\[0.4in]
  M_E &=& \left(\begin{array}{ccc} b_0 - 2 b_2 - 3(e_0 - 2 e_2) & 0 & 0\\
          0 & b_0 + b_1 + b_2 - 3(e_0 + e_1 + e_2) & 0\\
          0 & 0 & b_0 - b_1 + b_2 - 3(e_0 - e_1 + e_2)\\ 
          \end{array}\right),\\
  \label{eq:MNMECY}
\end{array}
\end{equation}

\noindent where the parameters have the values $a_0 = 18.935 + 0.000271681i,\ 
a_1 = -30.7989 + 0.0019887i,\ a_2 = 10.0361 + 0.00171701i,\ a_3 = -2.99072
-0.054757i,\ a_4 = 0.554859 - 0.234705i,\ a_5 = -0.066748 + 0.008155i,\ 
b_0 = 0.387756,\ b_1 = -0.539649,\ b_2 = 0.19327,\ d_0 = 8.60218,\ 
d_1 = -10.1912,\ d_2 = 3.72519,\ e_0 = -0.0227734,\ e_1 = 0.0228717,\ 
e_2 = -0.0115298$, all in GeV.  The common mass of the degenerate right-handed
neutrinos is determined to be $M_R = 2.4 \times 10^{12}$ GeV, so as to fit 
$\Delta m^2_{31} = 2.6 \times 10^{-3}\ {\rm eV^2}$ with the aid of the 
seesaw formula.  The authors find
\begin{equation}
  \begin{array}{rl}
      &m_1 = 7.7\ {\rm meV},\quad m_2 = 11.8\ {\rm meV},\quad m_3 = 50.7\ 
           {\rm meV},\\
      &\sin^2 2\theta_{23} = 1.00,\quad \sin^2 \theta_{12} = 0.29,\quad 
	 \sin^2 \theta_{13} = 0.0029.\\
  \end{array}
\label{eq:CYresults}
\end{equation}

\noindent  A value of $\tan \beta = 10$ is used to evolve the quark and charged 
lepton masses, though the neutrino masses and mixings have not been evolved 
downward to the electroweak scale.

\subsection{Dermisek - Raby $SO(10)$ Model}

This model is based on $SO(10)$ with a $D_3$ family symmetry \cite{DRmodel}.  
The charged lepton and down quark mass matrices have lopsided textures, but 
they are not so extreme as in the case of the Albright - Barr model:  
\begin{equation}
  M_N = \left(\begin{array}{ccc} 0 & \epsilon' \omega & -3\epsilon \xi \sigma\\
          - \epsilon' \omega & 3 \tilde{\epsilon}\omega & - 3\epsilon \sigma\\
	1.5 \epsilon \xi \omega & 1.5 \epsilon \omega & 1\\ 
          \end{array}\right) v\lambda \sin \beta, 
  \qquad M_E = \left(\begin{array}{ccc} 0 & \epsilon' & -3 \epsilon\xi\sigma\\
          - \epsilon' & 3 \tilde{\epsilon} & -3 \epsilon \sigma\\
          3 \epsilon \xi & 3 \epsilon & 1\\ 
          \end{array}\right) v\lambda \cos \beta,
  \label{eq:MNMEDR}
\end{equation}

\noindent where $v = 174$ GeV and the parameters have the values $\lambda = 0.64,\ 
\epsilon = 0.046,\ \sigma = 0.83 e^{0.618i},\ \tilde{\epsilon} = 
0.011 e^{0.411i},\ \rho = - 0.053 e^{0.767i},\ \epsilon' = - 0.0036,\ 
\xi = 0.12 e^{3.673i},\ \omega = 2\sigma/(2\sigma - 1)$.
The right-handed Majorana mass matrix is diagonal with mass eigenvalues
$M_1 = 1.1 \times 10^{10}$,\ $M_2 = - 9.3 \times 10^{11}$,\ 
$M_3 = 5.8 \times 10^{13}$ GeV. These parameters were chosen by making use 
of the central experimental values for $\Delta m^2_{21} = 7.9 \times 10^{-5}\ 
{\rm eV^2},\ \Delta m^2_{31} = 2.3 \times 10^{-3}\ {\rm eV^2},\ 
\sin^2 \theta_{12} = 0.295$ and $\sin^2 \theta_{23} = 0.51$ at the time of writing.
 The authors find
\begin{equation}
  \begin{array}{rl}
      &m_1 = 3.7\ {\rm meV},\quad m_2 = 9.6\ {\rm meV},\quad m_3 = 49.2\ 
           {\rm meV},\\
      &\sin^2 \theta_{13} = 0.0024.\\
  \end{array}
\label{eq:DRresults}
\end{equation}

\noindent   The evolution has been carried 
out with $\tan \beta = 49.98$ and $A_0 = - 6888.3$ GeV down to $M_Z$ and 
then down to the 1 GeV scale for the light neutrino masses.  

\subsection{Grimus - Kuhbock $SO(10)$ Model}

This model is based on $SO(10)$ with an $Z_2$ flavor symmetry \cite{GKmodel}.  
The Higgs fields appear in one each of the ${\bf 10}$, ${\bf 120}$, and 
$\overline{\bf 126}$ dimensional representations of $SO(10)$.  The fermion
mass matrices are generated by renormalizable Yukawa couplings of these 
Higgs fields to the three families placed in ${\bf 16}$'s.   
The Dirac neutrino and charged lepton mass matrices are written as linear
combinations of the Yukawa couplings of the ${\bf 10}$, ${\bf 120}$, and 
$\overline{\bf 126}$ Higgs representations, respectively, while the 
right-handed Majorana matrix is proportional only to the latter:
\begin{equation}
\begin{array}{rcl}
  M_N &=& r_H H' + r_D e^{i\psi_D}G' - 3r_F e^{i\zeta_u}F', \\
  M_E &=& H' + r_L e^{i\psi_L}G' - 3e^{i\zeta_d}F', \\
  M_R &=& rR^{-1} F', \\
\label{eq:MNMEGK}
\end{array}
\end{equation}

\noindent where the individual Higgs contributions to the mass matrices
are given by 
\begin{equation}
\begin{array}{rcl}
  H' &=& \left(\begin{array}{ccc} 0.716986 & 0 & 0\\
            0 & - 40.6278 & 0\\ 0 & 0 & 1114.41\\ \end{array}\right) 
            \times 10^{-3},\\
  G' &=& \left(\begin{array}{ccc} 0 & - 7.56737 & 0\\
            7.56737 & 0 & - 36.8224\\
	  0 & 36.8224 & 0\\ \end{array}\right) \times 10^{-3},\\
  F' &=& \left(\begin{array}{ccc} -0.0966851 & 0 & 4.25282\\
            0 & 12.3136 & 0\\
	  4.25282 & 0 & -61.6491\\ \end{array}\right) \times 10^{-3},\\
\end{array}
\end{equation}

\noindent  in the right-left order we have adopted.  The coefficient 
parameters are taken to be $r_H = 91.0759,\ r_F = 297.758,\ r_u = 7.14572,\ 
r_L = 1.33897,\ r_D = 3008.88,\ r_R = 2.90553 \times 10^{-17},\ 
\zeta_d = 19.66974^\circ,\ \zeta_u = -2.96594^\circ,\ \psi_L = 6.24258^\circ,
\ \psi_D = 179.85271^\circ$.   A value of $\tan \beta = 10$ is used to evolve 
the quark and charged lepton masses downward to the electroweak scale. One 
finds then that 
\begin{equation}
  \begin{array}{rl}
      &m_1 = 1.6\ {\rm meV},\quad m_2 = 9.2\ {\rm meV},\quad m_3 = 50.0\ 
           {\rm meV},\\
      &\sin^2 2\theta_{23} = 1.00,\quad \sin^2 \theta_{12} = 0.31,\quad 
	 \sin^2 \theta_{13} = 0.00059.\\
  \end{array}
\label{eq:GKresults}
\end{equation}

\noindent  Although the results are impressive, the authors do note that 
21 parameters have been introduced in order to obtain the results.  Hence
the model should be viewed as an existence proof that the lepton masses
and mixings as well as the quark masses and mixings can be described in
the framework of renormalizable couplings with only the three Higgs 
representations contributing to the Yukawa couplings.

\begin{table}[b]
\caption{\label{Table I} Higgs representations, flavor symmetries, and other 
noteworthy features of the five $SO(10)$ SUSY GUT models considered in 
this work.}
\vspace*{0.1in}
\begin{center}
\begin{tabular}{c|c|c|c|c|c|l}
\hline
Models & Higgs Content & Flavor Symmetry & $M_R$ (GeV) & $\tan \beta$ & 
     $\sin^2 \theta_{13}$ & \qquad\quad Interesting Features \\ \hline\hline
 AB & ${\bf 10},\ {\bf 16},\ \overline{\bf 16}, {\bf 45}$ & $U(1) \times Z_2 \times 
     Z_2$ & 
     $2.4\times 10^{14}$ & 5 & 0.0020 & Large $M_R$ hierarchy with lightest \\
     &  & & $4.5\times 10^8$ & & ($2.6^\circ$)  & two nearly degenerate leads to \\
     & & & $4.5\times 10^8$ & &  & resonant leptogenesis. \\[0.1in]
 CM & ${\bf 10},\ \overline{\bf 126}$ & $SU(2) \times (Z_2)^3$ &  $7.0\times 10^{12}$ 
     & 10 & 0.013 & Large $M_R$ hierarchy with heaviest  \\
     & & &  $4.5\times 10^9$ & &($6.5^\circ$)  & more than 3 orders of magnitude\\
     & & & $1.1\times 10^7$ & & & below GUT scale; large $\sin^2 \theta_{13}$. \\[0.1in]
 CY & ${\bf 10},\ \overline{\bf 126}$ & $S_4$ & $2.4\times 10^{12}$ & 
     10 & 0.0029 & Degenerate $M_R$ spectrum 4 orders\\
     & & & $2.4\times 10^{12}$ & & ($3.1^\circ$) & of magnitude  below GUT scale. \\
     & & & $2.4\times 10^{12}$ & & & \\[0.1in]
 DR & ${\bf 10},\ {\bf 45}$ & $D_3$ &  $5.8\times 10^{13}$ & 50 &0.0024 & Mild $M_R$ 
     hierarchy almost 3 orders\\
     & & &  $9.3\times 10^{11}$ & & ($2.8^\circ$) &   of magnitude  below GUT scale.\\
     & & & $1.1\times 10^{10}$ & &  &\\[0.1in]
 GK & ${\bf 10},\ {\bf 120},\ \overline{\bf 126}$ & $Z_2$ & $2.1\times 
     10^{15}$ & 10 & 0.00059 & Mild $M_R$ hierarchy just  1 order of\\
     & & &  $4.2\times 10^{14}$ & & ($1.4^\circ$) &  magnitude below GUT scale;\\
     & & & $6.7\times 10^{12}$ & & &rather small $\sin^2 \theta_{13}$.  \\[0.1in]
\hline
\end{tabular}
\end{center}
\end{table}

We close this section by summarizing the features of these models in Table I.
Three of the models have similar predictions for $\sin^2 \theta_{13} \sim 0.0025$, 
near the expected reach of the Double CHOOZ and Daya Bay reactor experiments 
\cite{newreactors}.  One of the models predicts a value near the present 
upper limit placed by the CHOOZ experiment \cite{CHOOZ}, while the fifth model predicts 
a value of $3 \times 10^{-4}$ which is essentially beyond reach until a 
Neutrino Factory becomes a reality.   We shall see what potential success
charged lepton flavor violation experiments will have in further distinguishing 
the viable models.

\section{LEPTON FLAVOR VIOLATION IN RADIATIVE DECAYS}

We now turn to the subject of charged lepton flavor violation that can occur in 
the radiative decays, $\mu \rightarrow e + \gamma$, $\tau \rightarrow \mu + 
\gamma$, and $\tau \rightarrow e + \gamma$.  In the SM with
the addition of three massive right-handed neutrinos, we observe that the 
individual lepton numbers, $L_e,\ L_\mu$ and $L_\tau$, are not individually
conserved.  In radiative lepton decays, the flavor violation arises in one 
loop, where the neutrino insertion involves lepton flavor-changing Yukawa
couplings of the left-handed and right-handed neutrinos; cf. Fig. 1.  The 
branching ratio BR21 defined by the ratio of the rate for the $\mu \rightarrow
e\gamma$ mode relative to the purely leptonic mode, $\mu \rightarrow 
\nu_\mu e \bar{\nu}_e$, is given by \cite{LeeShrock} 
\begin{equation}
\begin{array}{rcl}
      BR21 &=& \frac{3\alpha}{32\pi}\left|\sum_k U^*_{\mu k}\frac{m^2_k}{M^2_W}
	U_{ke}\right|^2\\
	&\simeq& \frac{3\alpha}{128\pi}\left(\frac{\Delta m^2_{21}}{M^2_W}\right)^2
		  	\sin^2 2\theta_{12} \sim 10^{-54}, \\
\end{array}
\label{eq:BR21SM}			
\end{equation}
where the $U$'s are elements of the PMNS mixing matrix.  Hence in the SM, the
expected branching ratio is immeasurably small, and the MEG experiment \cite{MEG}
looking for $\mu \rightarrow e\gamma$ would be expected to give a null result.
The present upper limit of $1.2 \times 10^{-11}$ was obtained 
by the MEGA collaboration \cite{MEGA}.

\begin{figure}
\includegraphics*[scale=0.9]{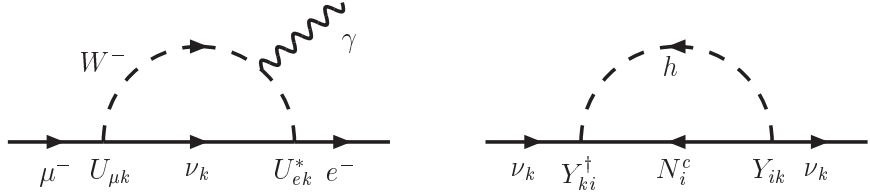}\\[-0.1in]
\caption{\label{Fig. 1.}  Example of a Feynman diagram for $\mu \rightarrow e 
\gamma$ with a neutrino mass insertion in the SM.}
\end{figure}

\indent In SUSY GUT models on the other hand, the leading log approximations involve
slepton-neutralino and sneutrino-chargino loops which contribute to the radiative 
lepton decays \cite{BorzMas}; cf. Fig.~2.  With more comparable heavy masses of the 
SUSY particles in the loops and lack of a Glashow-Iliopoulos-Maiani (GIM) mechanism, 
such a great suppression of the branching ratio does not occur.  We shall work in the 
CMSSM \cite{CMSSM}, where the soft-breaking scalar 
and gaugino  masses and trilinear scalar couplings are assumed to be universal at the 
GUT scale.  The lepton flavor violation then arises from evolution of the Yukawa 
couplings and soft-breaking parameters from the GUT scale down to the electroweak scale 
\cite{HMTY}.
 
Feynman diagrams for the LFV radiative decays in leading log approximation involve both 
neutralino - slepton ($\tilde{\chi}^0 - \tilde{\ell}$) loops and chargino - sneutrino 
($\tilde{\chi}^\pm - \tilde{\nu}$) loops with the emitted photon attached 
to the internal charged slepton or chargino, respectively.  Through evolution
from the GUT scale, the LFV neutrino Yukawa couplings are induced primarily in the mass 
squared submatrix for the $SU(2)_L$ doublet sleptons, $m^2_{\tilde{L}}(LL)$. We do not 
repeat the details here for this complicated calculation but rather refer the reader to 
the pioneering paper of Hisano, Moroi, Tobe, and Yamaguchi \cite{HMTY}.  We simply 
note that the radiative decay rate is given by 
\begin{equation}
  \Gamma(\ell^-_j \rightarrow \ell^-_i \gamma) = 
    \frac{e^2}{16\pi} m^5_{l_j}\left(|A^{(n)}_L + A^{(c)}_L|^2 + 
          |A^{(n)}_R + A^{(c)}_R|^2\right),
  \label{eq:rate}
\end{equation}

\noindent where $(n)$ and $(c)$ refer to the neutralino and chargino 
loop contributions to the transition form factors $A_L$ and $A_R$ connecting
leptons of opposite chirality. The branching ratio for the flavor-violating decay mode 
relative to the flavor-conserving purely lepton mode is then
\begin{equation}
  BR(\ell^-_j \rightarrow \ell^-_i \gamma) = \frac{48 \pi^3 \alpha}{G_F^2}
          (|A_L|^2 + |A_R|^2).
  \label{eq:BR}
\end{equation}

\begin{figure}
\includegraphics*[scale=0.9]{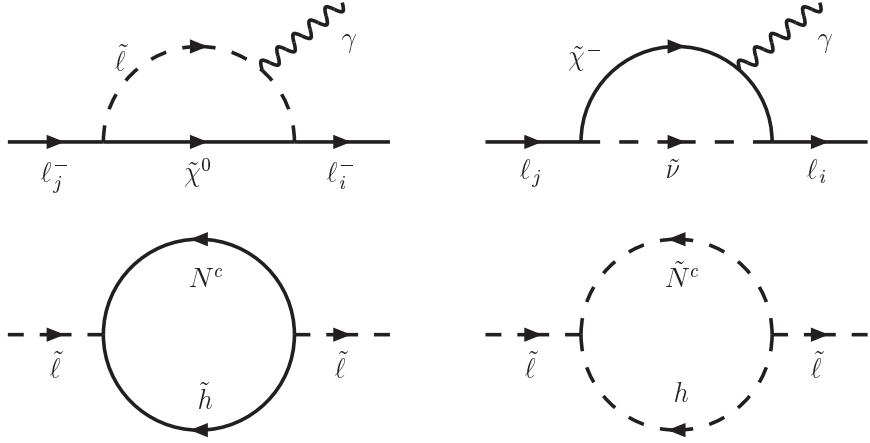}\\[-0.1in]
\caption{\label{Fig. 2.}  Examples of Feynman diagrams for slepton - neutralino 
 and sneutrino - chargino contributions to $\mu \rightarrow e \gamma$ in SUSY 
 models with slepton mass insertions.}
\end{figure}

In the leading log approximation with the largest contribution coming from the 
left-handed slepton mass matrix, the branching ratio is given by
\begin{equation}
 	BRji = \frac{\alpha^3}{G^2_F m^8_s}|(m^2_{LL})_{ji}|^2 \tan^2 \beta,
\label{eq:BR21SUSY}
\end{equation}
where
\begin{equation}
	(m^2_{LL})_{ji} = -\frac{1}{8\pi^2}m^2_0 (3+A^2_0/m^2_0)Y^\dagger_{jk}
		\log \left(\frac{M_G}{M_k}\right)Y_{ki}.
\label{eq:mLL}
\end{equation}
with the Yukawa couplings specified in the lepton flavor basis and the right-handed
Majorana matrix diagonal, so $M_k$ is just the kth heavy right-handed neutrino
mass, while $M_G$ is the GUT scale typically equal to $2 \times 10^{16}$ GeV and 
$m_s$ is some typical SUSY scalar mass.  Petcov and collaborators \cite{Petcov}
have shown that the full evolution effects as first calculated in \cite{HMTY} can be 
extremely well approximated by Eq. (\ref{eq:BR21SUSY}), if one sets 
\begin{equation}
	m^8_s \simeq 0.5 m^2_0 M^2_{1/2} (m^2_0 + 0.6 M^2_{1/2})^2.
\label{eq:m8approx}
\end{equation}
While the branching ratio for $\mu \rightarrow e + \gamma$ is well approximated by
the above equations, the branching ratios for $\tau \rightarrow \mu + \gamma$ and
$\tau \rightarrow e + \gamma$ must be scaled by the branching ratios for the 
flavor-conserving leptonic modes relative to their total decay rates.

We see that the MEG experiment \cite{MEG} has no chance of observing a positive 
signal in the SM for the $\mu \rightarrow e\gamma$ decay channel;  however, the
situation will be totally different for the SUSY GUT models we have chosen to 
consider.  In the CMSSM with universal soft-breaking parameters $m_0,\ M_{1/2}$
and $A_0$, for a given $\tan \beta$ and $sgn(\mu)$,
we consider the following correlations between the branching ratio and the soft-SUSY 
breaking parameters:\\[-0.3in]
\begin{itemize}
\item[(a)]  the branching ratio {\it vs.} $M_{1/2}$ for fixed $A_0 = 0$, for 
	example, with different choices of $m_0$;\\[-0.3in]
\item[(b)]  the allowed parameter space for $A_0/m_0$ {\it vs.} $M_{1/2}$, for specific 
	branching ratio ranges; \\[-0.3in]
\item[(c)]  the branching ratio for $\tau \rightarrow \mu \gamma$
	{\it vs.} that for $\mu \rightarrow e \gamma$, on a log-log plot, which are 
	related by
	\begin{equation}
        \begin{array}{rcl}
	\log {\rm BR}(\tau \rightarrow \mu \gamma) &=& \log {\rm BR}(\mu 
	\rightarrow e \gamma) + \log ({\rm BR32/BR21}) + \log {\rm BR}(\tau
        \rightarrow \nu_\tau \mu \bar{\nu}_\mu)\\
	&=& \log {\rm BR}(\mu \rightarrow e \gamma) + \log \left| 
	\frac{(Y^\dagger_\nu L Y_\nu)_{32}}{(Y^\dagger_\nu L Y_\nu)_{21}}\right|^2
	- 0.757,
\label{eq:BRratio}
\end{array}
\end{equation}

\noindent	where $L \equiv \log (M^2_G/M^2_R)$, and again the Yukawa matrices 
	are converted to the lepton flavor basis with $M_R$ diagonal.  Due to the 
	factorization of the soft-breaking parameters and the GUT model parameters
	in the approximate Eq. (\ref{eq:BR21SUSY}),  the slope is unity, and the 
	intercept is just the sum of the last two terms after correcting for the 
	$\tau \rightarrow \nu_\tau \mu \bar{\nu}_\mu$ branching ratio.   The length 
	of the straight line segment depends on the range of the soft-breaking 
	parameters chosen.  A similar plot can be made for the branching ratio for 
	$\tau \rightarrow e \gamma$ {\it vs.} that for $\mu \rightarrow  e \gamma$.
\end{itemize}	
					
For all three types of plots we have imposed the following soft 
parameter constraints \cite{PDB}:
\begin{equation}
\begin{array}{lrrcr}
    {\rm For}\ \tan \beta = 5, 10: &  m_0:  &   50 &\rightarrow& 400\ {\rm GeV}\\ 
	&  M_{1/2}: & 200 &\rightarrow& 1000\ {\rm GeV}\\
	&  A_0:  &  -4000 &\rightarrow& 4000\ {\rm GeV}\\
	{\rm For}\ \tan \beta = 50:	&  m_0:  &  500 &\rightarrow& 4000\ {\rm GeV}\\
	& M_{1/2}:  & 200 &\rightarrow& 1500\ {\rm GeV}\\
	&  A_0:  &  -50 &\rightarrow& 50\ {\rm TeV}\\
\end{array}
\label{eq:softlimits}
\end{equation}
In addition it is desirable to impose WMAP dark matter constraints in the 
neutralino, stau or stop coannihilation regions \cite{coannih}, where the lightest 
neutralino is the LSP.  These more restrictive constraints are well 
described by the quadratic polynomial for the soft scalar mass in terms 
of the soft gaugino mass \cite{WMAPconstraints}:
\begin{equation}
\begin{array}{rcl}
	m_0 &=& c_0 + c_1 M_{1/2} + c_2 M^2_{1/2},\\
	c_i &=& c_i(A_0,\ \tan \beta,\ sgn(\mu)).\\
\end{array}
\label{eq:DMlimits}
\end{equation}
\begin{figure}[b]
%\vspace*{-0.25in}
\hspace*{-0.6in} (a) \hspace*{2.9in} (b) \\
\includegraphics*[scale=0.6]{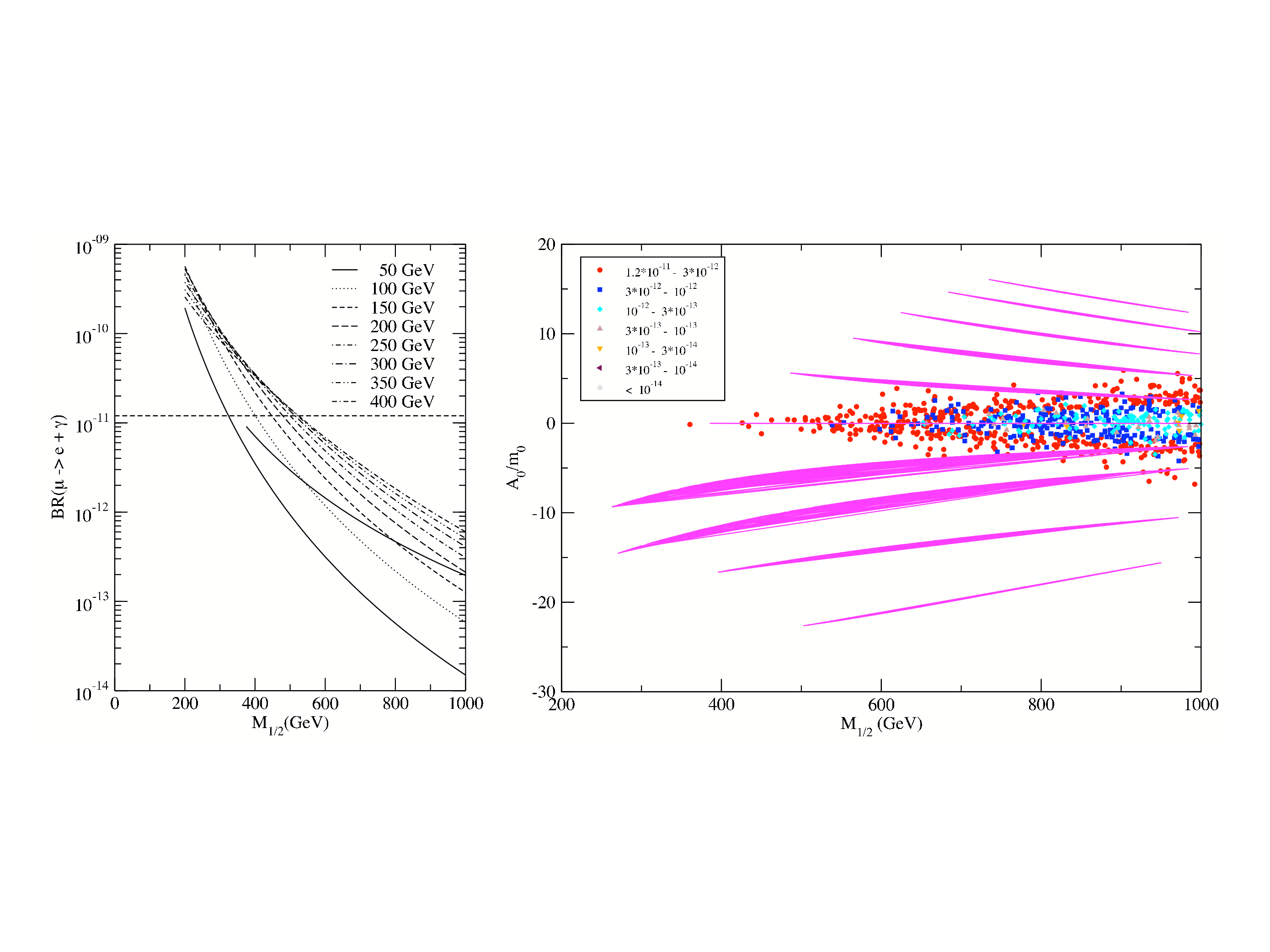}
\caption{\label{Fig. 3.}  Branching ratio predictions for $\mu \rightarrow e + 
  \gamma$ in the Albright - Barr model with $\tan \beta = 5$.     In (a) 
  $A_0$ is set equal to zero, while in (b) all three parameters, 
  $m_0,\ M_{1/2},\ A_0$, are allowed to vary.}
\end{figure}
where $m_0$ is bounded since $M_{1/2}$ is bounded.  If $M_{1/2}$ is too 
small, the present experimental bound on the Higgs mass \cite{PDB} of $m_h 
\gsim 114.4$ GeV may be violated or the neutralino relic density in the early 
universe will be too small, while if $M_{1/2}$ is too large the neutralino 
relic density will be too large.  We shall impose both limits on the 
DM constraint conditions for various values of $A_0$ and find that both lower 
and upper limits are placed on the branching ratios for each model.

\begin{figure}[t]
\vspace*{-0.25in}
\hspace*{-0.55in} (a) \hspace*{2.9in} (b) \\
\includegraphics*[scale=0.6]{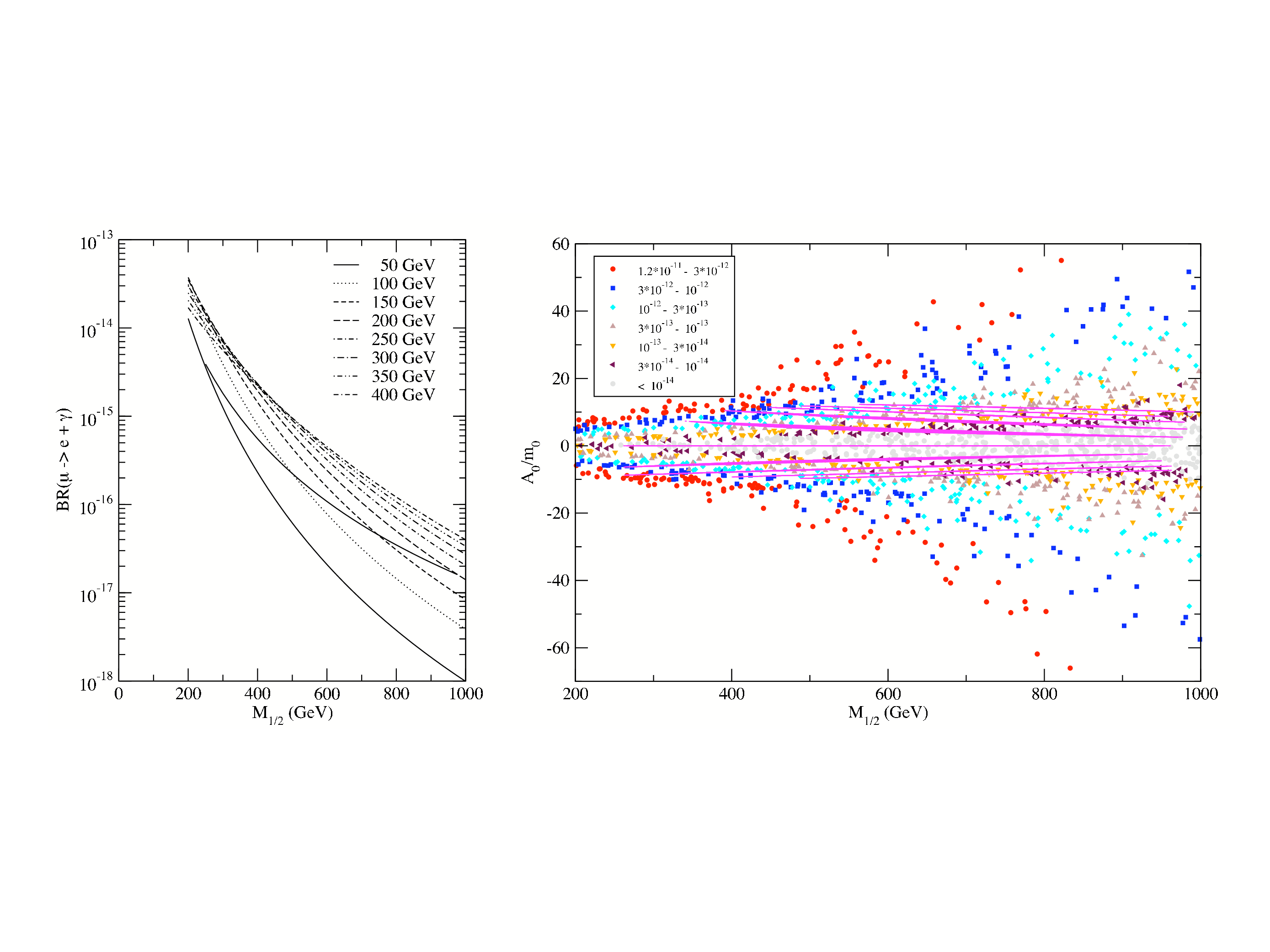}
 \caption{\label{Fig. 4.}  Branching ratio predictions for $\mu \rightarrow 
 e +  \gamma$ in the Chen - Mahanthappa model with $\tan \beta = 10$.}
 \end{figure}

\begin{figure}[b]
\hspace*{-0.55in} (a) \hspace*{2.9in} (b) \\
\includegraphics*[scale=0.6]{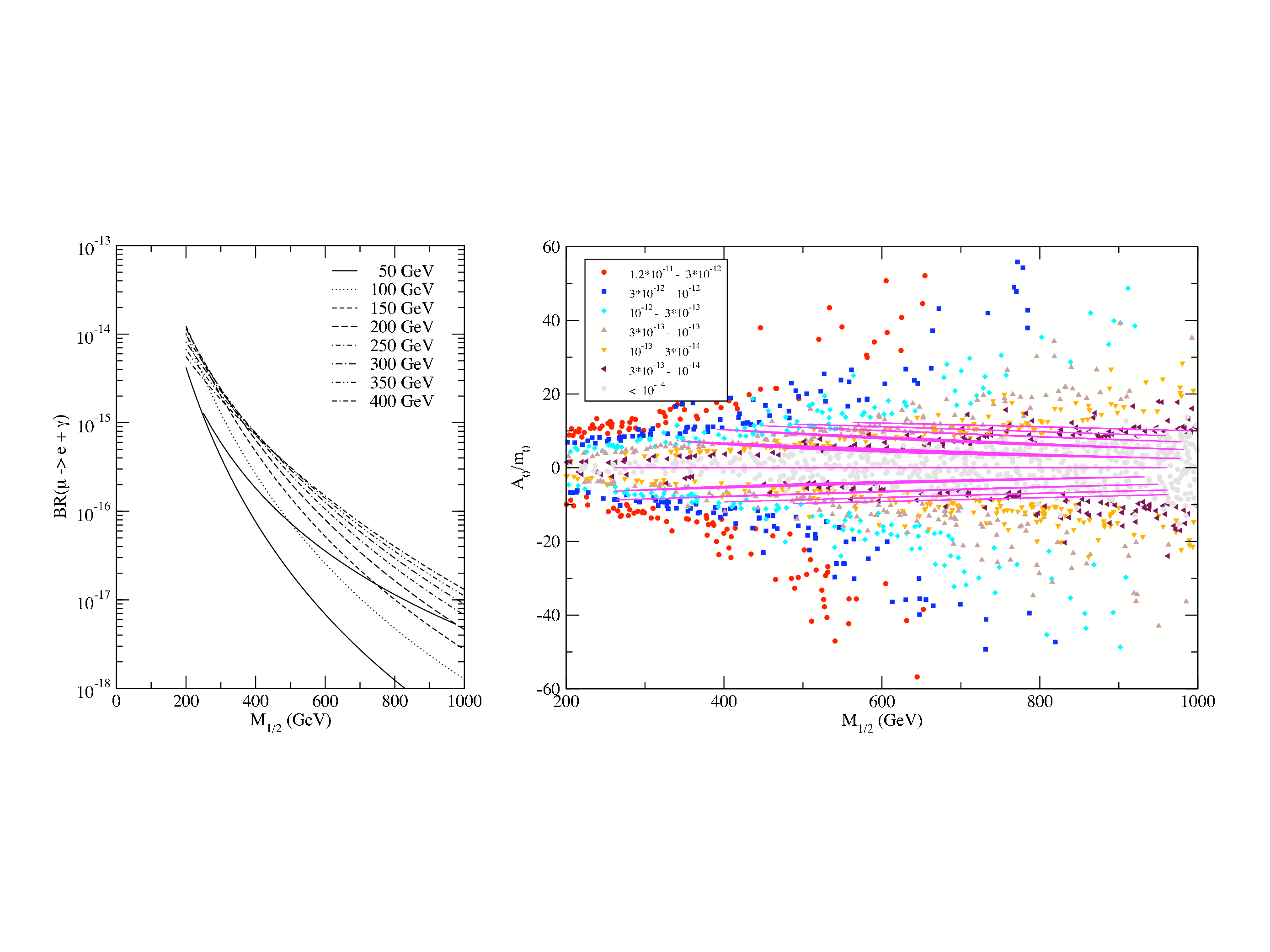}
\caption{\label{Fig. 5.}  Branching ratio predictions for $\mu \rightarrow 
 e + \gamma$ in the Cai - Yu model with $\tan \beta = 10$.}  
\end{figure}

For each model in Figs. 3(a), 4(a), 5(a), 6(a), and 7(a), we plot curves with constant 
values of $m_0$ indicated for the BR21 branching ratios as functions of $M_{1/2}$ with 
the trilinear coupling $A_0 = 0$ at the GUT scale.   The curve which cuts across 
the $m_0$ curves in each left-hand plot represents the WMAP dark matter constraint
for this value of $A_0$.  The horizontal broken line indicates the present 
experimental upper limit for this branching ratio \cite{MEGA}. 

In Figs. 3(b), 4(b), 5(b), 6(b), and 7(b), we allow $A_0$ to depart from zero and show 
scatterplots of $A_0/m_0$ {\it vs.}\ $M_{1/2}$.  The points are color-coded as indicated 
according to the branching ratio intervals in which they fall, with all points
below the present upper limit on the $\mu \rightarrow e \gamma$ branching
ratio.  The soft parameter constraints in Eq. (\ref{eq:softlimits}) have been 
imposed for the Monte Carlo selection of points.  The continuous curves represent
the WMAP dark matter constraints and are drawn in steps of 500 GeV from 
$A_0 = - 2.5$ TeV to $A_0 = 2.5$ TeV.

\begin{figure}[t]
\hspace*{-0.55in} (a) \hspace*{2.9in} (b) \\
\includegraphics*[scale=0.6]{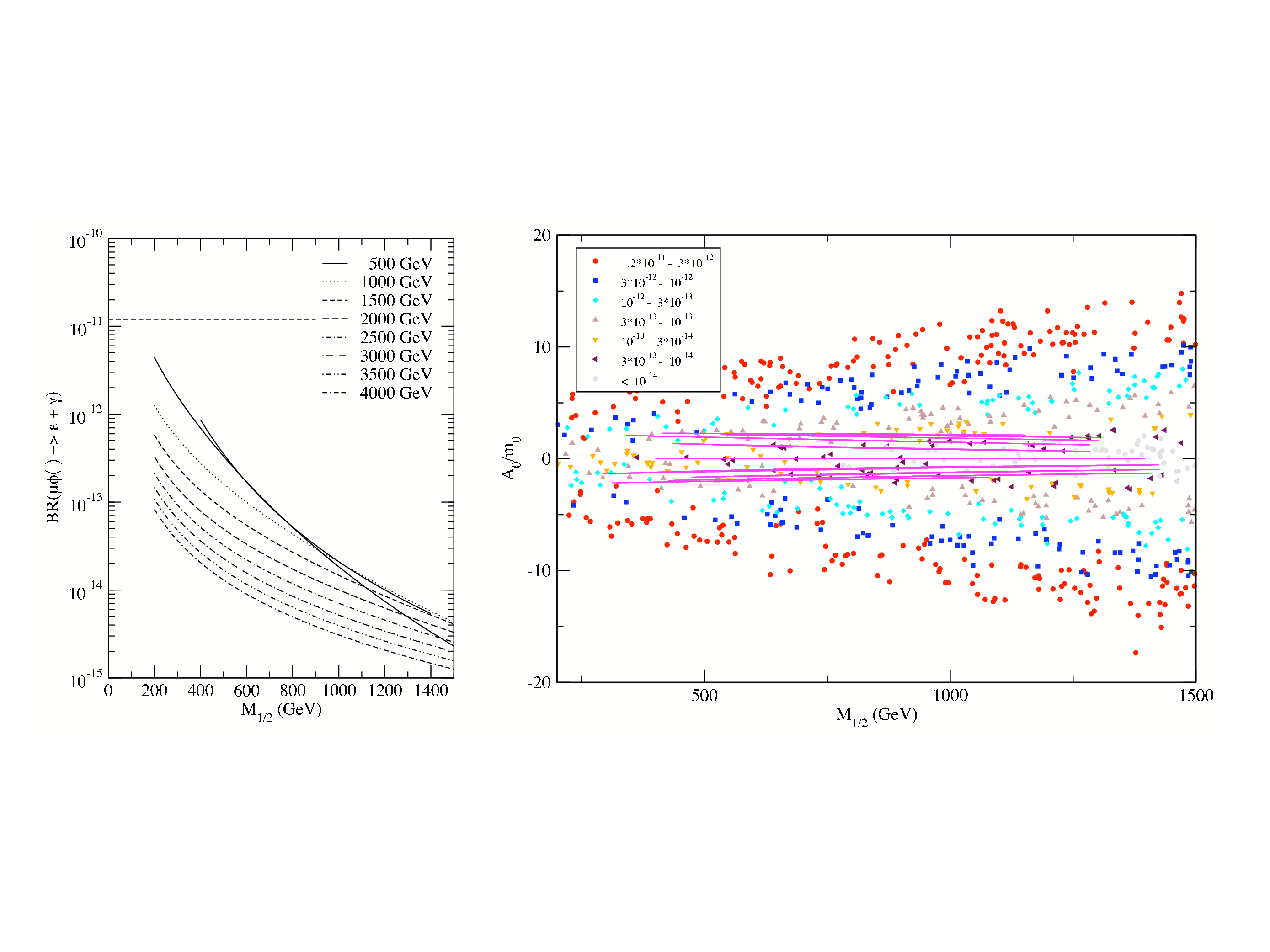}
\caption{\label{Fig. 6.}  Branching ratio predictions for $\mu \rightarrow e + 
\gamma$ in the Dermisek - Raby model with $\tan \beta = 50$.}
\end{figure}

\begin{figure}[b]
\hspace*{-0.55in} (a) \hspace*{2.9in} (b) \\
\includegraphics*[scale=0.6]{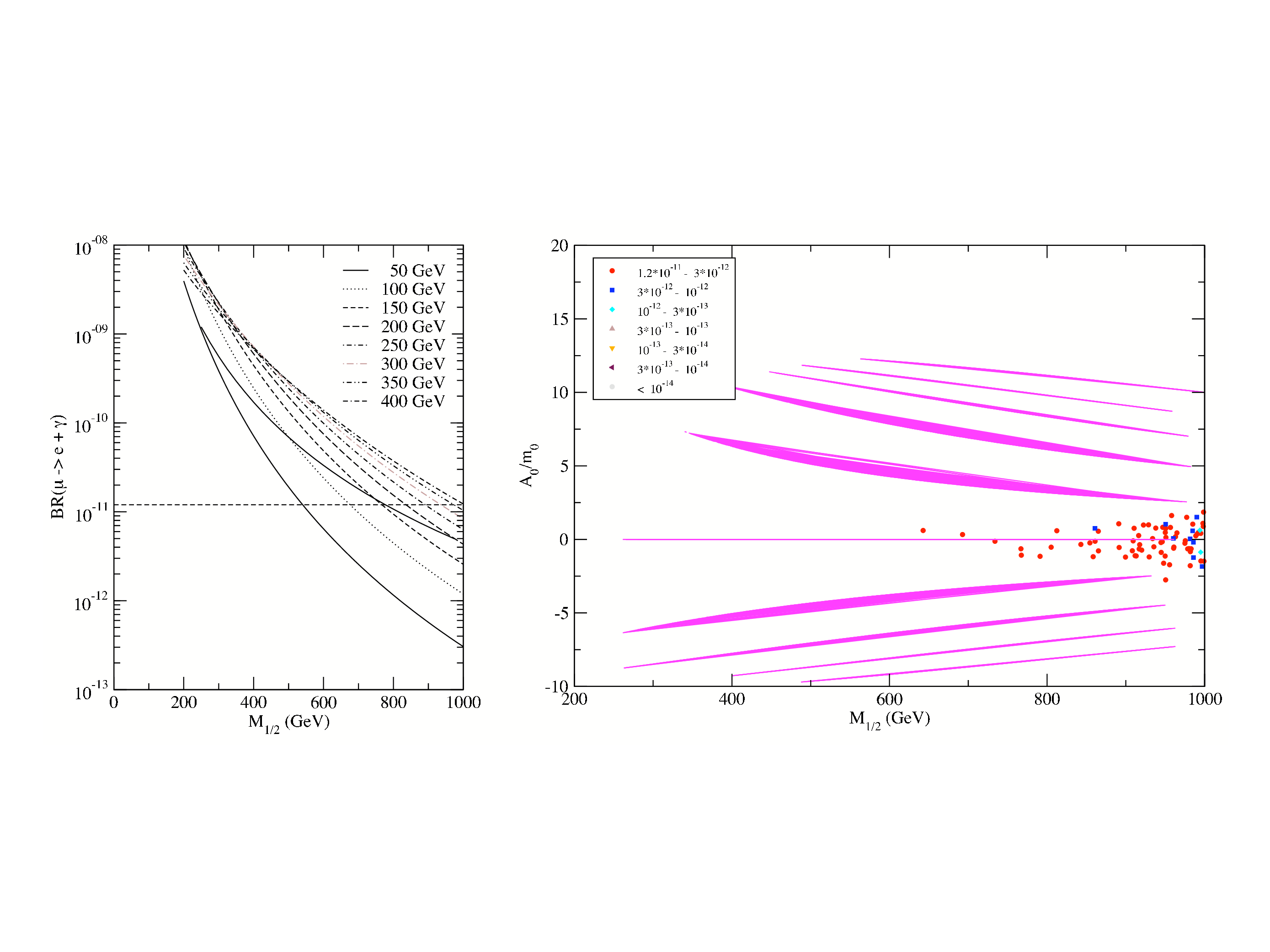}
\caption{\label{Fig. 7.}  Branching ratio predictions for $\mu \rightarrow e + 
   \gamma$ in the Grimus - K\"{u}bock model with $\tan \beta = 10$.}
\end{figure}
  
It is clear that the predicted branching ratio decreases as the universal soft gaugino 
mass at the GUT scale increases along the constant scalar mass curve.  The  
Grimus-Kuhbock (GK) and Albright-Barr (AB) models will be probed first and then the 
Dermisek-Raby (DR) model by the MEG experiment, while the other two models are 
essentially beyond reach, if $A_0 = 0$ as depicted.  One sees that higher values of 
BR21 are predicted for a given value of $M_{1/2}$ as $|A_0/m_0|$ increases.  For the 
AB and GK models the experimental branching ratio greatly limits the allowed ranges 
of $A_0/m_0$, while for the Chen-Mahanthappa (CM), Cai-Yu (CY), and DR models the 
dark matter constraints limit the allowed ranges of $A_0/m_0$.  In any case, the 
minimum predicted BR21 branching ratio occurs for $A_0 = 0$.

One can also present similar scatter plots for the $\tau \rightarrow \mu \gamma$
and $\tau \rightarrow e \gamma$ decay modes.  Since all three decay modes 
are intimately related in each model through the corresponding logarithmic 
terms such as that in Eq. (\ref{eq:BRratio}), the same scatter points will appear
with only the color-coding changed (assuming one imposes the BR21
experimental limit for each plot).  

%\vspace*{-0.2in}
\begin{figure}[b]
\includegraphics*[scale=0.6]{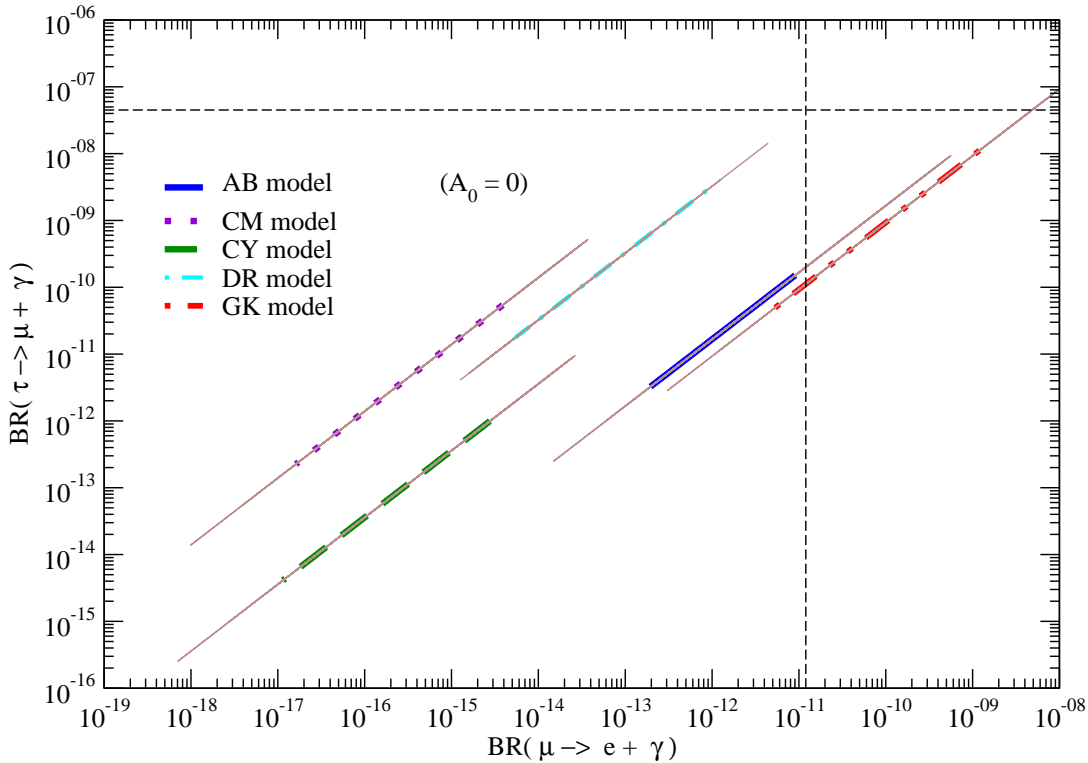}\\[-0.1in]
 \caption{\label{Fig. 8.}  Branching ratio predictions for $\tau \rightarrow \mu + 
 \gamma$ {\it vs.} branching ratio predictions for $\mu \rightarrow e + 
 \gamma$ in the five models considered.  The soft-SUSY breaking constraints
 imposed apply for the thin line segments, while the more restrictive WMAP 
 dark matter constraints apply for the thick line segments.   The present 
 experimental constraints are indicated by the dashed lines.}
 \end{figure}

\begin{figure}
\includegraphics*[scale=0.6]{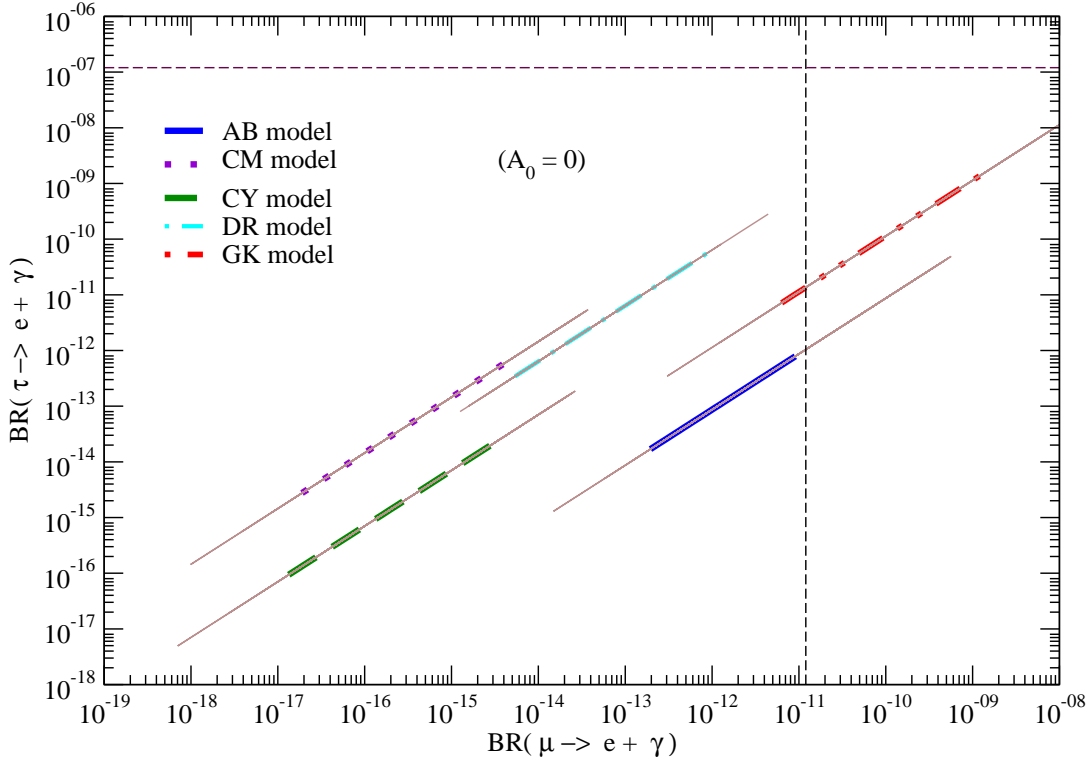}\\[-0.1in]
 \caption{\label{Fig. 9.}  Branching ratio predictions for $\tau \rightarrow e + 
 \gamma$ {\it vs.} branching ratio predictions for $\mu \rightarrow e + 
 \gamma$ in the five models considered.  The soft SUSY breaking constraints
 imposed apply for the thin line segments, while the more restrictive WMAP 
 dark matter constraints apply for the thick line segments.}
 \end{figure}

Instead, we present two log-log plots for the BR32 and BR31 branching ratios 
against that for BR21 in Figs. 8 and 9, where $A_0 = 0$ has again been imposed.  The 
thin line segments for each model observe the soft parameters constraints imposed, 
while the heavier line segments observe the more restrictive WMAP dark 
matter constraints.  The vertical dashed line reflects the present BR21 bound,
while the horizontal dashed line refers to the present BR32 or BR31 
experimental limit, respectively \cite{BRlimits}.  It is clear from these two plots 
that the ongoing MEG experiment stands the best chance of confirming the predictions 
for or eliminating the GK and AB models.  Even with a super-B factory \cite{superB}, 
the present experimental bounds on the BR32 and BR31 branching ratios can 
only be lowered by one or two orders of magnitude at most.

\begin{table}[b]
\caption{\label{Table II} Ratios of the branching ratios for the lepton flavor violating 
  $\tau$ decays and $\mu -e$ conversion on Ti relative to that for $\mu$ decay.}
\begin{center}
\begin{tabular}{c|c|c|c}
\hline
Models & BR($\tau \rightarrow \mu \gamma$)/BR($\mu \rightarrow e\gamma$) & 
	BR($\tau \rightarrow e \gamma$)/BR($\mu \rightarrow e\gamma$) & 	
	BR($\mu + Ti  \rightarrow e + Ti$)/BR($\mu \rightarrow e\gamma$) 
        \\ \hline\hline
AB 	& 16.7 & 0.09 & 0.33\\
CM   & $1.3 \times 10^4$ & 171 & 0.11 \\
CY    & 400 & 6.5 & 0.11 \\
DR   & $3.3 \times 10^3$ & 61.0 & 0.026 \\
GK   & 10.0 & 1.0 & 0.12 \\	
\hline
\end{tabular}
\end{center}
\end{table}

But recall that the line segments apply for the special case of $A_0 = 0$.
If one allows $A_0$ to depart from zero, the line segments will slide diagonally 
upward and toward the right along their presently depicted positions by amounts 
that can be estimated from Figs. 3 - 7.  Hence only the lower limits on the 
branching ratios are robust in Figs. 8, 9 and 11.  However, it is clear that ratios
of the branching ratios remain fixed for each model for any allowed $A_0/m_0$.
We present these ratios for the $\tau$ and $\mu - e$ conversion branching ratios 
relative to the $\mu \rightarrow e \gamma$ branching ratio in Table II.  The spread 
in numbers for the ratios in different models appears to be greater than that 
anticipated by the authors of Ref. \cite{vogel} for the class of models considered 
here.

\begin{table}[b]
\caption{\label{Table III} Summary of the relevant results for the five $SO(10)$ 
  SUSY GUT models considered in this work. The present experimental upper limits for 
  the branching ratios are indicated in the second line, while the third line of the 
  table gives the projected upper limit reaches for the Meg experiment, Super-B 
  factory, and next generation $\mu - e$ conversion experiment.}
\begin{center}
\begin{tabular}{c|c|c|c|c|r|r|r}
\hline
Models &  $\sin^2 \theta_{13}$ & $M_R$'s & $\tan \beta$ & 
     $|A_0/m_0|_{\rm max}$ & 
     \multicolumn{1}{c|}{BR21$(\mu \rightarrow e \gamma)$} & 
     \multicolumn{1}{c|}{BR32$(\tau \rightarrow \mu \gamma)$}  & 
     \multicolumn{1}{c}{BR($\mu + Ti \rightarrow e + Ti $)}\\ 
Expt.  & & (GeV) & & &  $ < 1.2 \times 10^{-11}$ & 
      $< 4.5 \times 10^{-8}$ &  $< 4 \times 10^{-12}$\\
Limits   & & & & & $ \rightarrow\ < 10^{-13}$ & 
      $\rightarrow\ < 10^{-9}$ &  $ \rightarrow\ < 10^{-18}$\\
     \hline\hline
 AB &  0.0020 & 
     $2.4\times 10^{14}$ & 5 & 5 &  $(0.2 - 9)\times 10^{-12}$ & $(0.03 - 1) 
     \times 10^{-10}$ & $(0.03 - 2) \times 10^{-12}$\\
      & ($2.6^\circ$) & $4.5\times 10^8$ & & & & \\
     & & $4.5\times 10^8$ & & & & \\[0.1in]
 CM &  0.013 & $7.0 \times 10^{12}$ & 
     10 & 12 & $(0.02 - 4)\times 10^{-15}$ & $(0.02 - 5)\times 10^{-11}$ &
     $(0.01 - 3)\times 10^{-16}$\\
      & ($6.5^\circ$) & $4.5\times 10^9$ & & & & \\
     & & $1.1\times 10^7$ & & & & \\[0.1in]
 CY &  0.0029 & $2.4 \times 10^{12}$ & 
     10 & 12 & $(0.02 - 5)\times 10^{-15}$ & $(0.04 - 9)\times 10^{-13}$ &
     $(0.03 - 6)\times 10^{-16}$\\
      & ($3.1^\circ$) & $2.4\times 10^{12}$ & & & & \\
     & & $2.4\times 10^{12}$ & & & & \\[0.1in]
 DR &  0.0024 & $5.8\times 10^{13}$ & 50 & 2.5 & 
     $(0.05 - 8)\times 10^{-13}$ & $(0.02 - 3)\times 10^{-9}$ &
     $(0.01 - 2)\times 10^{-14}$\\
     & ($2.8^\circ$) & $9.3\times 10^{11}$ & & & & \\
      & & $1.1\times 10^{10}$ & & & & \\[0.1in]
 GK &  0.00059  & $2.1\times 10^{15}$ & 10 & 2 &  $(0.4 - 80)\times 10^{-11}$ &  
     $(0.004 - 1)\times 10^{-8}$ &  $(0.02 - 5)\times 10^{-11}$ \\
      & ($1.4^\circ$) & $4.2\times 10^{14}$ & & & & \\
     & & $6.7\times 10^{12}$ & & & & \\[0.1in]
\hline
\end{tabular}
\end{center}
\end{table}

In Table III we summarize the relevant findings from our study of the five models.  
The branching ratio ranges apply for the $A_0 = 0$ case and with the stricter WMAP 
dark matter constraints imposed.  It is clear that the five predictive $SO(10)$ 
SUSY GUT models considered have very representative right-handed neutrino mass spectra 
and predictions for $\sin^2 \theta_{13}$. The CM, DR, and GK models have massive 
hierarchical spectra with $M_3$ ranging from $10^{13}$ to $10^{15}$ GeV. The CY model,
on the other hand, has a degenerate spectrum with $M_R \sim 3 \times 10^{12}$ GeV, 
while the AB model has degenerate $M_1$ and $M_2$ which can lead to resonant 
leptogenesis.  The CM model has a relatively large $\sin^2 \theta_{13}$ prediction
which should be observable at the upcoming reactor neutrino experiments, Double CHOOZ
and Daya Bay.  The AB, CY and DR models have similar predictions for 
$\sin^2 \theta_{13}$ which will make observation of $\bar{\nu}_e \rightarrow 
\bar{\nu}_\mu$ oscillation somewhat marginal at those reactors and in the proposed 
NO$\nu$A and T2K long-baseline experiments \cite{novat2k} without
a SuperBeam source.  For the GK model, the observation of such a low 
$\sin^2 \theta_{13}$ prediction would only take place with a Neutrino Factory.
But it is clear from Table III and the previous figures that the GK and AB models 
will be tested first with the MEG experiment.  From our discussion it is then clear
that the LFV branching ratios are more sensitive to large $M_3$ than to large 
$\theta_{13}$ in the models considered.  In previous generic studies of SUSY GUT
models, the rare branching ratios were nearly equally sensitive to each of the 
two parameters \cite{genericresults}.

\section{LEPTON FLAVOR VIOLATION IN $\mu - e$ CONVERSION}

It is also of interest to consider lepton flavor violation in the $\mu$ to $e$ 
conversion process in Titanium, $\mu + Ti \rightarrow e + Ti$.
While there is no such ongoing experiment, preliminary discussion is underway 
to propose one which will lower the present limit.  We shall see that the 
conversion rate predictions relative to the muon capture process, $\mu + Ti 
\rightarrow \nu_\mu + Sc$, are such that all five models considered in this paper 
can potentially be eliminated, if no signal is observed. 

While predictions for the conversion process in the SM are infinitesimally small, 
large enhancements in the CMSSM framework again occur by virtue of massive super
partners appearing in loop diagrams involving $\gamma,\ Z$ and Higgs penguins
and boxes.   Of these, Arganda, Herrero and Teixeira have shown that the $\gamma$ 
penguin contributions dominate the others by at least two orders of magnitude 
\cite{AHT}. The slepton-neutralino and sneutrino-chargino loop contributions to the 
$\gamma$ penguins are shown in Fig. 10.  The virtual massive $N^c$ and $\tilde{N}^c$ 
with their Yukawa couplings again appear in the slepton loops along with a higgsino
or Higgs particle, respectively.

\begin{figure}[h]
\includegraphics*[scale=0.8]{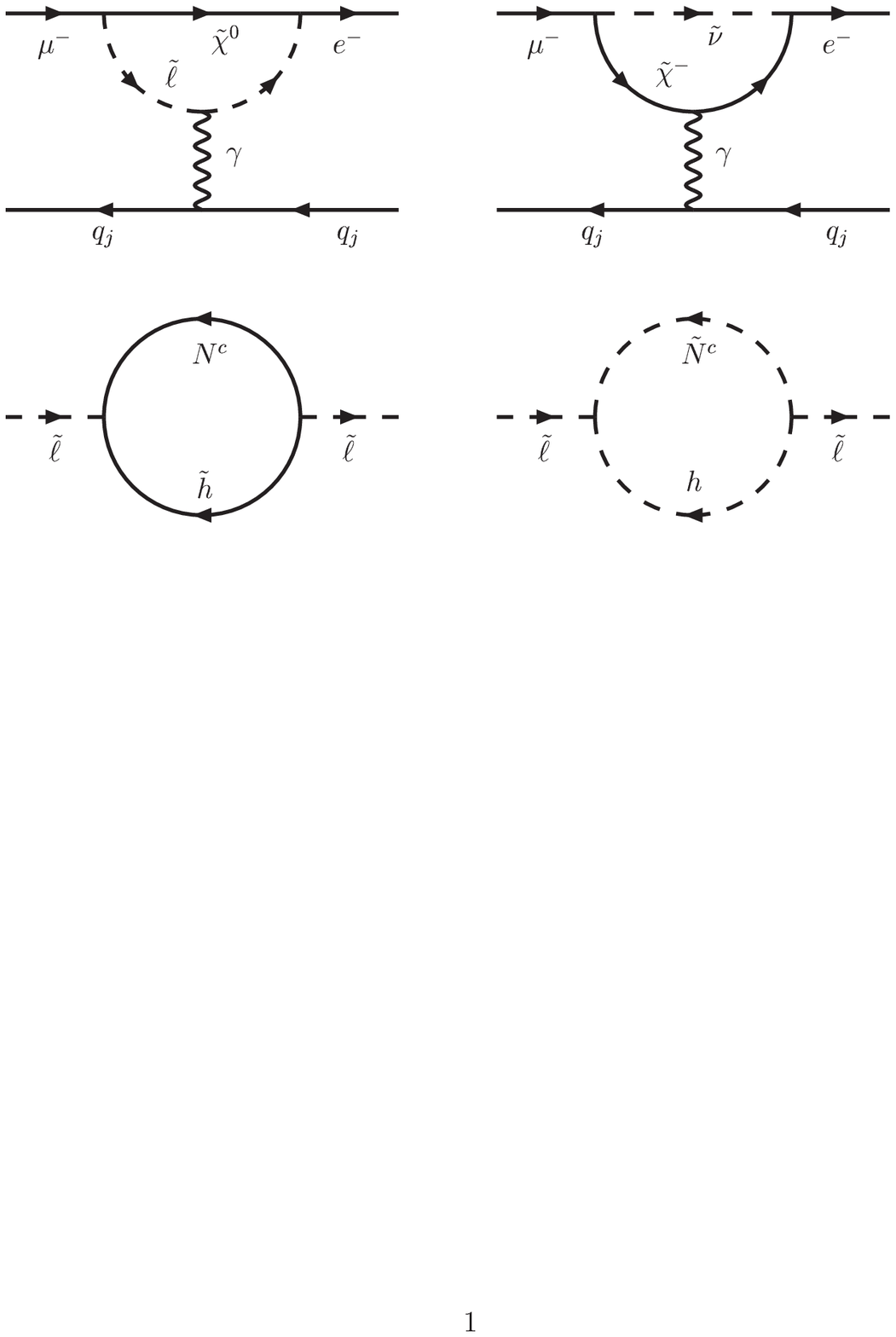}\\[-0.2in]
\caption{\label{Fig. 10.}  Examples of Feynman diagrams for slepton - neutralino 
and sneutrino - chargino contributions to $\mu - e$ conversion in SUSY models
with slepton mass insertions.}
\end{figure}

The complex formulas for lepton flavor violation in this process were first derived by 
Hisano, Moroi, Tobe, and Yamaguchi \cite{HMTY}.  If we restrict our attention to the 
$\gamma$ penguin contribution as suggested by Arganda et al. \cite{AHT}, one finds 
the $\mu - e$ conversion rate is given by 
\begin{equation}
	\Gamma(\mu \rightarrow e) = 4\alpha^5 Z^4_{\rm eff}Z m^5_\mu |F(q)|^2
		\left[ |A^L_1 - A^R_2|^2 + |A^R_1 - A^L_2|^2 \right],
\label{eq:mueconvrate}
\end{equation}

\noindent where here $A^{L,R}_1$ are form factors for the vertices connecting
leptons of equal chirality, while $A^{L,R}_2$ are combinations of the electric
dipole and magnetic dipole transition form factors previously denoted by $A_{L,R}$.
For a $^{48}_{22}Ti$ target, $Z_{\rm eff} = 17.6$ and the nuclear form factor is 
$F(q^2 \simeq -m^2_\mu) \simeq 0.54$ \cite{HMTY}.  In the case of the conversion 
process, we have explicitly carried out the full evolution running from the GUT scale to 
the $Z$ scale.  The $\mu - e$ conversion branching ratio is then obtained from
the conversion rate above by scaling it with the $\mu$ capture rate on $Ti$, 
which is quoted in \cite{mueconv} as $(2.590 \pm 0.012) \times 10^6\ {\rm sec^{-1}}$
with the present experimental limit on the conversion branching ratio found to be 
$R \leq 4 \times 10^{-12}$.

\begin{figure}[t]
\vspace*{-0in}
\includegraphics*[scale=0.6]{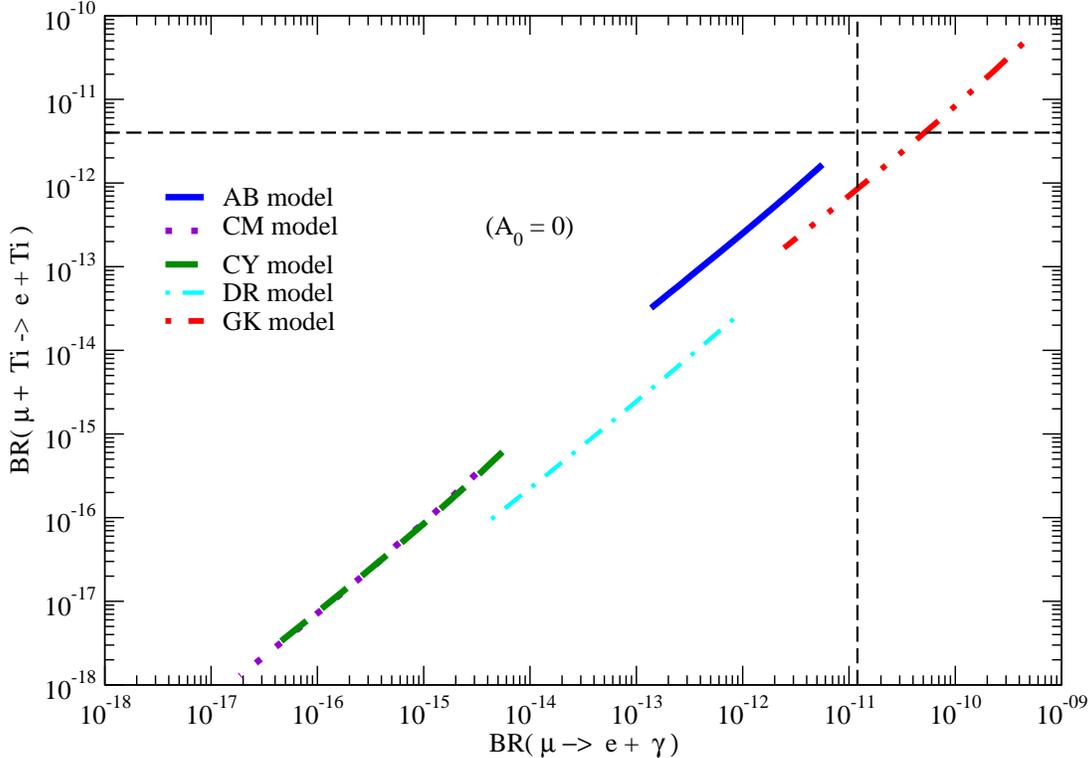}\\[-0.2in]
 \caption{\label{Fig. 11.}  Branching ratio predictions for $\mu - e$ conversion
 {\it vs.} branching ratio predictions for $\mu \rightarrow e + \gamma$ in the 
 five models considered.  Only thick line segments are shown reflecting the 
 application of the more restrictive WMAP dark matter constraints.  Note that the
 predictions for the CM and CY models overlap.}
 \end{figure}

In Fig. 11.  we show a plot of the $\mu - e$ conversion branching ratio {\it vs.} 
the $\mu \rightarrow e\gamma$ branching ratio for each of the five models considered.  
We have limited the line segments by applying the WMAP dark matter constraints of 
Sec. III. It is clear that the GK and AB models would be tested first, followed by 
the DR, CY and CM models.  In fact, a first generation $\mu - e$ conversion 
experiment may be able to reach a branching ratio of $10^{-17}$, while a second 
generation experiment may lower the limit from the present value down to $10^{-18}$
\cite{mueprojection}. If such proves to be the case and no signal is seen, all five 
models will be eliminated.  Hence the conversion experiment is inherently more 
powerful than the MEG experiment looking for $\mu \rightarrow e \gamma$ which is 
designed to reach a level of $10^{-13} - 10^{-14}$, sufficient only to eliminate 
the GK and AB models.  The caveat, of course, is that MEG is now starting to take 
data, while no new conversion experiment has been approved to date.

\section{Conclusions}

There have been many theoretical models constructed which aim to explain the neutrino 
masses and mixing patterns. While the predictions for the value of $\theta_{13}$ in  
these models may provide a way to distinguish some of these models, as shown in the 
survey over 63 models we have carried out in Ref. \cite{Albright:2006cw}, the rare 
LFV processes, such as $\mu \rightarrow e + \gamma$ and $\mu-e$ conversion can provide 
an even more sensitive way to disentangle these models. We have investigated these rare 
processes in five SUSY SO(10) models that are currently still viable and highly 
predictive, making use of the allowed parameter space for the soft SUSY breaking 
parameters in the CMSSM framework. Utilizing the WMAP dark matter constraints, lower 
bounds on the branching ratios of these rare processes can be placed, and we find that 
at least three of the five models considered give rise to prediction for $\mu 
\rightarrow e + \gamma$ that will be tested at MEG. More interestingly, the next  
generation $\mu-e$ conversion experiment should be sensitive to the predictions of 
all five models, making it an even more robust way to test these models.
While generic studies have emphasized
the important dependence of the branching ratios on the reactor neutrino angle, 
$\theta_{13}$, and the mass of the heaviest right-handed neutrino,  
we find the latter to be by far the more significant in the models tested.

\begin{center}
  {\bf ACKNOWLEDGMENTS}
\end{center}

We thank Stephen Martin for making available to us his evolution program for the 
Yukawa couplings and soft SUSY breaking parameters. The work of M.-C.C. is 
supported, in part, by the National Science Foundation under Grant No. PHY-0709742.  
One of us (C.H.A.) thanks the members of the Theory Group at Fermilab for 
their kind hospitality. Fermilab is operated by the Fermi Research Alliance under 
contract No. DE-AC02-07CH11359 with the U.S. Department of Energy.

\vspace*{0.2in}
\noindent {\it Note added.}--After completion of this work, the authors of Ref. 
\cite{muecorrection} informed
us that a suppression factor should be applied to the lepton flavor violating
branching ratios calculated in our paper.  This suppression arises from a QED
radiative correction to the effective dipole operators responsible for the rare
decays.  The multiplicative factor in question is given by $1 - \delta$, where 
$\delta = (8\alpha/\pi)\log(\Lambda/m_j)$, $\Lambda \sim 250 - 1000$ GeV is 
the sparticle mass scale responsible for the LFV, and $m_j = m_\mu\ {\rm or}\ 
m_\tau$ for the rare radiative muon or tau decays.  For $\mu \rightarrow e\gamma$,
the $\delta$ correction amouts to roughly 0.15, so the branching ratio is about 
0.85 times the rate given in the paper.  The authors thank Andrzej Czarnecki 
and Ernest Jankowski for pointing out their earlier work to them. 

%\newpage


\begin{thebibliography}{99}

\bibitem{atm}
        Y. Ashie {\it et al.} (Super-Kamiokande Collaboration), Phys. Rev. D {\bf 71}, 
	112005 (2005);
	J. Hosaka {\it et al.} (Super-Kamiokande Collaboration), Phys. Rev. D {\bf 73}, 
	112001 (2006);
	B. Aharmin {\it et al.} (SNO Collaboration), Phys.Rev. C {\bf 72}, 055502 (2005);
\bibitem{reactor}
        M. Apollonio {\it et al.} (CHOOZ Collaboration), Phys. Lett.B {\bf 420}, 
	397 (1998); ibid., Eur. Phys. J C {\bf 27}, 331 (2003);
	F. Boehm {\it et al.} (Palo Verde Collaboration), Phys.Rev. D {\bf 64}, 
	112001(2001);
	T. Araki {\it et al.} (KamLAND Collaboration), Phys.Rev.Lett.{\bf 94}, 081801 
	(2005);
	S. Abe {\it et al.} (KamLAND Collaboration), arXiv:0801.4589.
\bibitem{lbl}
        M. Ahn {\it et al.} (K2K Collaboration), Phys. Rev. D {\bf 74}, 072003 (2006);
	D.G. Michael {\it et al.} (MINOS Collaboration), Phys. Rev. Lett. {\bf 97},
	191801 (2006); (MINOS Collaboration), arXiv:0708.1495.
\bibitem{Maltoni:2004ei} 	
	M. Maltoni, T. Schwetz, M.A. Tortola and J.W.F. Valle, New J. Phys. {\bf 6}, 
	122 (2004).
\bibitem{Albright:2006cw} 
	C.H. Albright and M.-C. Chen, Phys. Rev. D {\bf 74}, 113006 (2006).
\bibitem{BorzMas}
	F. Borzumati and A. Masiero, Phys. Rev. Lett. {\bf 57}, 961 (1986).
\bibitem{Chen:2003zv}	
	For recent reviews, see, e.g. M.-C.~Chen and K.~T.~Mahanthappa,
  	Int.\ J.\ Mod.\ Phys.\ A {\bf 18}, 5819 (2003); AIP Conf.\ Proc.\  {\bf 721}, 
	269 (2004);
	G. Altarelli and F. Feruglio, New J. Phys. {\bf 6}, 106 (2004); 
	S.F. King, Rept. Prog. Phys. {\bf 67}, 107 (2004); 
	Z.z. Xing, Int. J. Mod. Phys. A {\bf 19}, 1 (2004); 
	R.N.Mohapatra, New J. Phys. {\bf 6}, 82 (2004); 
	A. Yu. Smirnov, Int. J. Mod. Phys. A {\bf 19} 1180 (2004); 
	R.N. Mohapatra and A.Y. Smirnov, Ann. Rev. of Nucl. and Part. Sci., {\bf 56}, 
	569 (2006);
	R.N. Mohapatra {\it et al.}, Rep. Prog. Phys. {\bf 70}, 1757 (2007);  
	A. Strumia and F. Vissani, arXiv:hep-ph/0606054.
\bibitem{generic}
          A. Masiero, S.K. Vempati, and O. Vives, Nucl. Phys. {\bf B649}, 189 (2003);
	New J. Phys. {\bf 6}, 202 (2004);
	F. Deppisch, H. P\"{a}s, A. Redelbach, R. R\"{u}ckl, and Y. Shimizu,
	Eur. Phys. J. C {\bf 28}, 365 (2003); 
	F. Deppisch, H. P\"{a}s, R. R\"{u}ckl, and A. Redelbach, Phys. Rev.D {\bf 73},
	033004 (2006);
	E. Arganda and M.J. Herrero, Phys. Rev. D {\bf 73}, 055003 (2006);
	L. Calibbi, A. Faccia, A. Masiero, and S.K. Vempati, J. High Energy Phys. 
	{\bf 7}, 012 (2007).
\bibitem{CasasIbarra}
        J.A. Casas and A. Ibarra, Nucl. Phys. {\bf B618}, 171 (2001).
\bibitem{exceptions}
        A. Kageyama, S. Kaneko, N. Shimoyama, and M. Tanimoto, Phys. Lett. B 
	{\bf 527}, 206 (2002);
        Q. Shafi and Z. Tavartkiladze, Nucl. Phys. {\bf B772}, 133 (2007); {\bf B778},
	216(E) (2007);
	S. Antusch and S.F. King, Phys. Lett. B {\bf 659}, 640 (2008).
	An earlier test of one of the models considered in this work for the 
	$\mu \rightarrow e \gamma$ branching ratio was published in 
	E. Jankowski and D.W. Maybury, Phys. Rev. D {\bf 70}, 035004 (2004).
\bibitem{seesaw}
	P. Minkowski, Phys. Lett. B {\bf 67}, 421 (1977); 
  	T. Yanagida, in {\it  Proceedings of the Workshop on the Unified Theory and 
    	Baryon Number in the Universe, Tsukuba, Japan 1979}, edited by O. Sawada 
    	and A. Sugamoto (KEK,  Tsukuba, 1979), p. 95; 
  	M. Gell-Mann, P. Ramond, and R. Slansky, in {\it Supergravity}, edited by 
  	P. van Nieuwenhuizen and D. Z. Fredman (North-Holland, Amsterdam, 
  	1979), p. 315; 
 	S.L. Glashow, in {\it Proceedings of the 1979 Cargese Summer Institute
    	on Quarks and Leptons}, edited by M. Levy, J.-L. Basdevant, D. Speiser, 
	J. Weyers, R. Gastmans, and M. Jacob (Plenum Press, New York, 1980), p. 687;
  	R.N. Mohapatra and G. Senjanovic, Phys. Rev. Lett. {\bf 44}, 912 (1980). 
\bibitem{PMNS}
	B.~Pontecorvo, Zh. Eksp. Teor. Fiz. {\bf 33}, 549 (1957) [Sov. Phys. JETP 
	{\bf 6}, 429 (1957)];
	Z.~Maki, M.~Nakagawa and S.~Sakata, Prog. Theor. Phys. {\bf 28}, 870 (1962).
\bibitem{PDB}
 	Review of Particle Physics, W.-M. Yao et al., J. Phys. G {\bf 33}, 1 (2006).
\bibitem{minimal}
	K.S. Babu and R.N. Mohapatra, Phys. Rev. Lett. {\bf 70}, 2845 (1993).
\bibitem{lopsided}
	K.S. Babu and S.M. Barr, Phys. Lett. B {\bf 381}, 202 (1996);
	C.H. Albright, K.S. Babu, and S.M. Barr, Phys. Rev. Lett. {\bf 81}, 1167 (1998).
\bibitem{normal}
	C.H. Albright, Phys. Lett. B {\bf 599}, 285 (2004).
\bibitem{ABmodel}
	C.H. Albright and S.M. Barr, Phys.Rev. D {\bf 64}, 073010 (2001).
\bibitem{Amodel}
	C.H. Albright, Phys. Rev. D {\bf 72}, 013001 (2005); {\bf 74}, 039903(E) (2006).
\bibitem{leptogen}
	A. Pilaftsis and T.E.J. Underwood, Nucl. Phys. {\bf B692}, 303 (2004).
\bibitem{CMmodel}
	M.-C. Chen and K.T. Mahanthappa, Phys. Rev. D {\bf 70}, 113013 (2004).
\bibitem{CYmodel}
	Y. Cai and H.-B. Yu, Phys. Rev. D {\bf 74}, 115005 (2006).
\bibitem{DRmodel}
	R. Dermisek and S. Raby, Phys.Lett. B {\bf 622}, 327 (2005).
\bibitem{GKmodel}
	W. Grimus and H. K\"{u}hbock, Phys. Lett. B {\bf 643}, 182 (2006).
\bibitem{newreactors}
	S.A. Dazley (Double CHOOZ Collaboration), in Proceedings of NuFACT05, Nucl. 
	Phys. B, Proc. Suppl. {\bf 155}, 231 (2006); 
	J. Cao (Daya Bay Collaboration), ibid. {\bf 155}, 229 (2006).
\bibitem{CHOOZ} 
	M. Apollonio {\it et al.} (CHOOZ Collaboration), Phys. Lett. B {\bf 420}, 
	397 (1998); Eur. Phys. J. C {\bf 27}, 331 (2003);
  	cf. also F. Boehm {\it et al.} (Palo Verde Collaboration), Phys.Rev. D {\bf 64}, 
	112001(2001).
\bibitem{LeeShrock}
	B.W. Lee and R.E. Shrock, Phys. Rev. D {\bf 16}, 1444 (1977).
\bibitem{MEG}
	A. Baldini, Nucl. Phys. B, Proc. Suppl. {\bf 168}, 334 (2007).	
\bibitem{MEGA}
	M.L. Brooks {\it et al.} (MEGA Collaboration), Phys. Rev. Lett. {\bf 83}, 1521 
	(1999).
\bibitem{CMSSM}
        For references to the CMSSM, cf. K.A. Olive, submitted to the SUSY07
	proceedings arXiv:0709.3303.
\bibitem{HMTY}
	J. Hisano, T. Moroi, K. Tobe, and M. Yamaguchi, Phys. Rev. {\bf 53}, 2442 (1996).
\bibitem{Petcov}
	S.T. Petcov, S. Profumo, Y. Takanishi, and C.E. Yaguna, Nucl. Phys. {\bf B676}, 
	453 (2004).
\bibitem{WMAP}
	D.N. Spergel {\it et al.} (WMAP Collaboration), Astrophys. J. Suppl. Ser. 
	{\bf 170}, 377 (2007).
\bibitem{coannih}
	P. Binetruy, G. Girardi, and P. Salati, Nucl. Phys. {\bf B237}, 285 (1984);
	K. Griest and D. Seckel, Phys. Rev. D {\bf 43}, 3191 (1991).
\bibitem{WMAPconstraints}
	L.S. Stark, P. H\"{a}fliger, A. Biland, and F. Pauss, J. High Energy Phys. 
	{\bf 0508}, 059 (2005).
\bibitem{BRlimits}
        B. Aubert {\it et al.} (BABAR Collaboration), Phys. Rev. Lett. {\bf 96}, 041801 
	(2006); {\bf 95}, 041802 (2005); 
	K. Abe {\it et al.} (Belle Collaboration), in {\it Proceedings of the 
	ICHEP06 Conference, Moscow, 2006}.
\bibitem{superB}
        A.G. Akeroyd {\it et al.} (SuperKEKB Physics Working Group), 
	arXiv:hep-ex/0406071.
\bibitem{vogel}
	V. Cirigliano, A. Kurylov, M.J. Ramsey-Musolf, and P. Vogel, Phys. Rev. Lett. 
	{\bf 93}, 231802 (2004).
\bibitem{novat2k}
        Y. Hayato {\it et al.} (T2K Collaboration), Nucl. Phys. B, Proc. Suppl. 
	{\bf 143}, 269 (2005); 
        D.S. Ayres {\it et al.}, Technical Design Report for the NO{$\nu$}A Experiment 
	E929 at Fermilab, 2007.
\bibitem{genericresults}
        S. Antusch, E. Arganda, M.J. Herrero, and A.M. Teixeira, J. High Energy Phys. 
	11, 090 (2006), cf. Ref. \cite{generic}.
\bibitem{AHT}
        E. Arganda, M.J. Herrero, and A.M. Teixeira, J. High Energy Phys. 10, 104 (2007).
\bibitem{mueconv}
        C. Dohmen {\it et al.} (SINDRUM II Collaboration), Phys. Lett. B {\bf 317}, 
	631 (1993).
\bibitem{mueprojection}
        W. Molzon, in {\it Fermilab Program Planning Report Presented before the P5 
	Committee, January 2008}.
\bibitem{muecorrection}
        A. Czarnecki and E. Jankowski, Phys. Rev. D {\bf 65}, 113004 (2002).
\end{thebibliography}
\end{document}